\documentclass[twocolumn]{aastex6}

\begin{document}


\title{GASP I: Gas stripping phenomena in galaxies with MUSE}


\author{Bianca M. Poggianti$^1$, Alessia
  Moretti$^1$, Marco Gullieuszik$^1$, Jacopo
  Fritz$^2$, Yara Jaff\'e$^3$, Daniela
  Bettoni$^1$, Giovanni Fasano$^1$, Callum 
  Bellhouse$^4,3$, George Hau$^3$, Benedetta
  Vulcani$^{5,1}$, Andrea Biviano$^6$, Alessandro Omizzolo$^7$, Angela Paccagnella$^8,1$,
  Mauro D'Onofrio$^8$, Antonio Cava$^9$, Y.-K. Sheen$^{10}$, Warrick Couch$^{11}$, Matt Owers$^{12,11}$}
\affil{$^1$INAF-Astronomical Observatory of Padova   
vicolo dell'Osservatorio 5   
35122 Padova, Italy, 
$^2$Instituto de Radioastronomia y Astrofisica, UNAM, Campus 
  Morelia, A.P. 3-72, C.P. 58089, Mexico, 
$^3$European Southern Observatory, Alonso de Cordova 3107,
  Vitacura, Casilla 19001, Santiago de Chile, Chile, 
$^4$University of Birmingham School of Physics and Astronomy,
  Edgbaston, Birmingham, England, 
$^5$School of Physics, The University of Melbourne, Swanston St \&
  Tin Alley Parkville, VIC 3010, Australia, 
$^6$INAF-Osservatorio Astronomico di Trieste, via G.B. Tiepolo 11,
  34131 Trieste, Italy, 
$^7$Vatican Observatory, Vatican City State, 
$^8$Department of Physics and Astronomy, University of Padova,
  vicolo dell'Osservatorio 5, 35122 Padova, Italy, 
$^9$Observatoire de Geneve, University of Geneve, 51 Ch. des
  Maillettes, 1290 Versoix, Switzerland, 
$^{10} $Korea Astronomy and Space Science Institute, Daejeon, 305-348, Korea, 
$^{11}$Australian Astronomical Observatory, North Ryde, NSW 1670, Australia,
$^{12}$Department of Physics and Astronomy, Macquarie University, NSW
  2109, Australia}





\begin{abstract}

GASP (GAs Stripping Phenomena in galaxies with MUSE) is a new
integral-field spectroscopic survey with MUSE at the VLT
aiming at studying 
gas removal processes in galaxies.
We present an overview of the survey and show a first example
of a galaxy undergoing strong gas stripping. 
GASP is obtaining deep MUSE data for 114 galaxies at
z=0.04-0.07 with stellar masses in the range
$10^{9.2}$-$10^{11.5} M_{\odot}$ in different environments (galaxy
clusters and groups, over more than four orders of magnitude in halo
mass). GASP targets galaxies with 
optical signatures of unilateral debris or tails
reminiscent of gas stripping processes (``jellyfish galaxies''), 
as well as a control sample of disk galaxies with no morphological anomalies.
GASP is the only existing Integral Field Unit (IFU) survey covering both the main galaxy
body and the outskirts and surroundings, where the IFU data can
reveal the presence and the origin of the outer gas.
To demonstrate GASP's ability to probe the physics of gas and stars, 
we show the complete analysis of a textbook case of a ``jellyfish'' galaxy, JO206. This is a massive
galaxy ($9 \times 10^{10} M_{\odot}$) in a low-mass cluster ($\sigma
\sim 500 \rm \, km \, s^{-1}$), at a small projected clustercentric
radius and a high relative velocity, with $\geq$90kpc-long tentacles of
ionized gas stripped away by ram pressure. We present the spatially resolved
kinematics and physical properties of gas and stars, and depict the
evolutionary history of this galaxy.

\end{abstract}




\section{Introduction} 

How gas flows in and out of galaxies is one of the central questions in galaxy 
formation and evolution. 
In the current hierarchical paradigm, the hot gas in dark matter haloes cools, 
feeding the interstellar medium present in the galaxy disk and
replenishing the cold gas stock that is needed to form new stars
(White \& Rees 1978, Efstathiou \& Silk 1983). 
Any process that prevents the gas from cooling efficiently, or 
removes gas either from the halo or from the disk, has fundamental 
consequences for the subsequent galaxy 
history. 

Gas-averse processes abound. 
Virial shock heating of circumgalactic gas is an obvious contender. 
According to hydro-cosmological simulations, above a critical halo 
mass of $\sim 10^{12} M_{\odot}$, the radiative 
cooling rate is not sufficient to prevent a stable virial shock, while 
cold gas streams can substain the gas inflow 
onto massive haloes only at $z>1-2$ (Birnboim \& Dekel 
2003, Kere\v{s} et al. 2005, Dekel \& Birnboim 2006, Dekel et 
al. 2009). Thus, haloes with $M> 10^{12} M_{\odot}$ at $z<1$ should be 
naturally deprived of their gas supply by virial shocks. 

Circumgalactic gas might be prevented from cooling also by 
simply removing the hot gas halo in the so-called ``strangulation'' scenario 
(Larson et al. 1980, Balogh et al. 2000). Since the halo gas 
is more loosely bound to the galaxy than the disk gas, it can be more 
easily stripped either by ram pressure stripping or by tidal effects once a 
galaxy is accreted onto a more massive halo. 

Both processes mentioned above leave intact the gas that is already in the disk. 
Several other mechanisms can instead affect the disk gas in a direct 
way. Their origin can be 
internal to galaxies themselves, such as galactic winds due to star 
formation or an active galactic nucleus (Veilleux et al. 2005, King \&
Pounds 2015, Erb 2015), or external (Boselli \&
Gavazzi 2006). Among the latter, there is ram pressure stripping due 
to the pressure exerted by the intergalactic medium (Gunn \& Gott 1972),
thermal evaporation (Cowie \& Songaila 1977) and turbulent/viscous 
stripping (Nulsen 1982). All of these affect the gas, but not directly 
the stellar component of galaxies. Tidal mechanisms instead affect 
both gas and stars, and include 
strong galaxy interactions and mergers 
(Barnes \& Hernquist 1992), tidal effects of a cluster as a whole (Byrd \& Valtonen 1990) and the so-called 
``harassment'', that is the cumulative effect of several weak and fast 
tidal encounters (Moore et al. 1996). 

While the cosmic web of gaseous filaments expected to feed galaxies
can be observed (e.g. Cantalupo et al. 2012, 2014),
pure-gas accretion onto galaxies is very difficult to probe observationally, 
and direct observational evidence is still rare (e.g. Sancisi et
al. 2008, Bouch\'e et al. 2013),
except in specific cases like X-ray cooling flows (Peterson \& Fabian,
2006).

Direct observations of gas flowing out of galaxies are relatively
easier, though a complete picture of how and why galaxies lose gas is still
far from being reached. 
Many studies lack the
multiwavelength data required 
to know the fate of the different gas phases (molecular gas, neutral and 
ionized hydrogen, and X-ray gas), but detailed observations 
are beginning to accumulate for a few galaxies (e.g. Sun et al. 2010, 
Vollmer et al. 2012, Yagi et al. 2013, J\'achym et al. 2013, Abramson
et al. 2011, 2016).

The survey presented in this paper focuses on those processes that
affect the gas in the disk, and not the stellar component.
The most convincing body of evidence for gas-only removal
comes from observations of internally-driven outflows and ram pressure
stripping. 

Quasar-driven and starburst-driven massive outflows of the cold phase and the
ionized phase are 
now observed both at low- and high-z, though what fraction of the outflowing gas
rains back onto the galaxy is still an open question (Feruglio et al. 2010, Steidel et
al. 2010, Fabian
2012, Bolatto et al. 2013, Genzel et al. 2014, Cicone et al. 2014,
Wagg et al. 2014, 
Cresci et al. 2015).

HI studies have convincingly shown the efficiency of ram pressure
stripping of the neutral gas in nearby galaxy clusters (e.g. Haynes et
al. 1984, Cayatte et al. 1990, Kenney et al. 2004, Chung et al. 2009,
Vollmer et al. 2010, Jaff\'e et al. 2015),
and sometimes also in groups (e.g. Rasmussen et al. 2006, 2008,
Verdes-Montenegro et al. 2001, Hess et al. 2013).
Conclusions from molecular studies are more debated,  
but overall the molecular gas seems to be removed in clusters, though  
less efficiently than the atomic gas (cf. Kenney \& Young 1989 and Boselli et al. 1997,
Boselli et al. 2014).  
Ionized gas studies based on $\rm H\alpha$ imaging are another
excellent tracer of gas stripping in clusters (e.g. Gavazzi et
al. 2002, Yagi et al. 2010, Yoshida et al. 2012, Fossati et al. 2012, Boselli et al. 2016).
Even more powerful are IFU studies, that are able to reveal 
the gas that is stripped and ionized and also provide
the kinematical and physical properties of the ionized gas as well as the stars
(Merluzzi et al. 2013, 2016, Fumagalli et al. 2014, Fossati et
al. 2016). Hydrodynamical simulations of ram pressure stripping
  describe the formation of these gas tails and their evolution
(Abadi et al. 1999, Quilis et al. 2000, Roediger \& Br\"{u}ggen 2007, Kapferer et al. 2008,
Tonnesen \& Bryan 2012, Tonnesen \& Stone 2014, Roediger et al. 2014, see reviews by Roediger
2009 and Vollmer 2013). 

Notably, stars are often formed in the stripped gas (e.g. Kenney \&
Koopmann 1999, Yoshida et
al. 2008, Smith et al. 2010, Hester et al. 2010, Kenney et
al. 2014, J\'achym et al. 2014).
Galaxies in which stars are born within the stripped gas tails can therefore be
identified also from ultraviolet or blue images, in which the newly
born stars produce a recognizable signature
(Cortese et al. 2007, Smith et al. 2010, Owers et al. 2012).
The most striking examples of this are the so-called ``jellyfish
galaxies''\footnote{To our knowledge, the first work using the term
  ``jellyfish'' was Smith et al. (2010).}, that exhibit tentacles of material that appear to be
stripped from the galaxy body, making the galaxy resemble animal jellyfishes
(Fumagalli et al. 2014, Ebeling et al. 2014, Rawle et al. 2014).
In the last years, the first optical systematic searches for gas stripping candidates
have been conducted (Poggianti et al. 2016, McPartland et al. 2016).

GASP\footnote{http://web.oapd.inaf.it/gasp/index.html} is a new integral-field spectroscopy survey with MUSE aiming at
studying gas removal processes from galaxies.
It is observing 114 disk
galaxies at z=0.04-0.07 comprising both a sample with optical signatures of
unilateral debris/disturbed morphology, suggestive of gas-only removal
processes, and a control sample lacking such signatures. Galaxies with
obvious tidal features/mergers were purposely excluded. GASP is thus
tailored for investigating those processes that can remove gas, and
only gas, from the disk, though we cannot exclude that tidal effects
are partly or fully responsible for the morphologies observed in some
of the targets. The
GASP data themselves will clarify the physical causes of the gas
displacement. Being based on optical spectroscopy, this study 
can reveal the ionized gas component. Neutral and molecular
studies of the GASP sample are ongoing, as described in \S8.

The most salient characteristics of GASP are the following:

1) {\it Galaxy areal coverage.}
In addition to the galaxy main body, the IFU data cover the
  galaxy outskirts, surroundings and eventual tails, out to $\sim
  50-100$ kpc away from the main galaxy component, corresponding to
$> 10 R_e$. The galaxy outskirts and surroundings are crucial for detecting the 
extraplanar gas and eventual stars. 
The combination of large field-of-view (1'$\times$1') and sensitivity of MUSE
at the GASP redshifts allows us to observe galaxies out to 
large radii, while maintaining a good spatial resolution ($\sim 1$
kpc). This is a unique feature of GASP, as other large IFU surveys
typically reach out to $2.5-3 R_e$ at most (see Table~3 in Bundy et
al. 2015).

2) {\it Environment.}
One of the main goals of GASP is to study gas removal processes as a
function of environment, and understand in what environmental conditions are such processes efficient.
GASP explores a wide range of environments, from galaxy clusters
to groups and poor groups. Its targets are located in dark matter
haloes with masses spanning four orders of magnitude ($10^{11}-10^{15}
M_{\odot}$). 

3) {\it Galaxy mass range.}
GASP galaxies have a broad range of stellar masses
($10^{9.2}-10^{11.5}$), therefore it is possible to study the efficiency
of gas removal processes and their effects on the star formation
activity as a function of galaxy mass and size.

In this first paper of the series we present the GASP scientific goals (\S2), describe the survey (\S3), the
observations (\S4) and
the analysis techniques (\S6) and we show the results for a 
strongly ram-pressure-stripped massive galaxy in a low-mass cluster
(\S7). The current status and data release policy are described in \S5.
For the first results of the GASP survey, 
readers are refereed also to other papers of the first
set (Bellhouse et al. Paper II, Fritz et al. Paper III, Moretti et al. Paper IV and Gullieuszik et
al. Paper V).


In all papers of this series we adopt a standard concordance cosmology with
$H_0 = 70 \, \rm km \, s^{-1} \, Mpc^{-1}$, ${\Omega}_M=0.3$
and ${\Omega}_{\Lambda}=0.7$ and a Chabrier (2003) IMF.

\section{Scientific drivers} 

The key science questions to be addressed with GASP are the following:

1. Where, why and how is gas removed from galaxies? (\S2.1)

2. What are the effects of gas removal on the star formation activity
and on galaxy quenching? (\S2.2)

3. What is the interplay between the gas physical conditions and the
activity of the galaxy central black hole? (\S2.3)

4. What is the stellar and metallicity history of galaxies prior to
and in absence of gas removal? (\S2.4)


\subsection{The physics of gas removal}

GASP seeks the physical mechanism responsible for the gas removal. For
each GASP galaxy, we address the following questions:
Is gas being removed? By which physical process (ram
pressure stripping, tidal effects, AGN etc)? What is the
amount and fraction of gas that is being removed?
GASP will evaluate this comparing the morphology and kinematics of the
stellar and gaseous components of each galaxy using the stellar
continuum and the emission lines in the MUSE spectra, respectively,
and measuring gas masses from $\rm H\alpha$ fluxes as described in \S6.3.

Whatever the gas removal process at work, the GASP data can shed light on the
rich physics involved, observing how the gas removal proceeds and
what are the timescales involved, how the
kinematics and morphology of the gas are affected, whether there are
large scale outflows,
what are the metallicities and the dust content of
the gas, and which is the cause for gas ionization (star
formation, shocks, AGN). Most of these quantities can
be directly measured from the MUSE spectra, either
from individual lines (gas kinematics and morphology, outflows, see \S6.2)  
or from the emission line ratios (metallicity, dust, ionization mechanism, see
\S6.3).

The general questions we wish to investigate with the complete
GASP sample want to shed light on fundamental issues
regarding the loss of gas from galaxies, such as:
For which fraction of galaxies are gas-only removal processes relevant?
For which types/masses of galaxies? 
In which environments? Only in clusters, or also in groups? 
Where in clusters (for which clustercentric distances/velocities/orbits
etc)? And, is the efficiency of gas removal enhanced during halo-halo
merging?

Ram pressure stripping calculations are obtained both with 
analytical methods and hydrodynamical simulations (e.g. Gunn \& Gott 1972,
Jaffe' et al. 2015, and references above). They predict how the
efficiency of gas stripping depends on the galaxy and environmental 
parameters under certain assumptions, which can be tested
with the GASP sample. Moreover, there is some observational 
evidence that the efficiency of gas stripping is 
  enhanced by shocks and strong gradients in the X-ray ICM 
(Owers et al. 2012, Vijayaraghavan \& Ricker 2013), but only a large 
sample such as GASP can inequivocally determine a correlation and 
the necessary physical conditions.
Finally, the first cosmological hydrodynamical simulations including the effects of ram pressure,
tidal stripping and satellite-satellite encounters on the HI gas in 
different environments make predictions on the relative roles of the
various mechanisms as a function of halo mass and redshift
(e.g. Marasco et al. 2016). 
Studies like GASP are the natural observational counterparts to
corroborate or reject the theoretical predictions.

\subsection{Gas, star formation and quenching}


Overall, the star formation activity in galaxies has strongly declined 
since $z \sim 2$, for the combination of two effects: a large number 
of previously star-forming galaxies have evolved into passive 
(i.e. has stopped forming stars), and the star formation rate in still 
star-forming galaxies has, on average, decreased (Guglielmo et 
al. 2015). Innumerable observational evidences point to this,
including the evolution of the passive fraction with time and the 
evolution of the star formation rate--stellar mass relation (e.g. Bell et
al. 2004, Noeske et al. 2007). 
On a cosmic 
scale, this leads to a drop in the star formation rate density of the 
Universe (Madau \& Dickinson 2014). 

One of the most debated questions is what drives the star 
formation decline. The availability of gas is central for this problem. 
The simplest explanation is that 
galaxies ``run out of gas'': they are deprived of gas repenishment due to 
virial shocks or strangulation, and consume the disk gas for star formation. 
In alternative, or in addition, they can 
have their star formation shut off 
by one or 
more of the internal or external physical processes acting on the disk gas, 
described in the previous section. 

GASP 
provides the spatially resolved ongoing star formation activity
and star formation history. 
Thus, the GASP data allow us to link the gas removal process with its effects on
the galaxy stellar history, determine whether
the star formation activity is globally enhanced, or suppressed due
to the mechanism at work, and
how the quenching of star formation proceeds within the galaxy
and on what timescale. The goal is to understand
how many stars are formed in the stripped gas, how the extraplanar
star formation contributes to the intergalactic medium and, more in general,
what is the impact of gas-only removal
processes for galaxy quenching.


\subsection{Gas and AGN}
While supermassive black holes are thought to be ubiquituous at least
in massive galaxies, an Active Galactic Nucleus (AGN) powered by accretion of matter onto the
black hole is much rarer (Kormendy \& Ho 2013, Brinchmann et al. 2004). Several candidates have been proposed as
``feeding mechanisms'' able to trigger the AGN activity. 
These include all those processes
that can cause large scale gas inflow in the galaxy central regions,
such as gravitational torques due to galaxy mergers or interactions
(Di Matteo et al. 2005, Hopkins et al. 2006), or disk instability due for example
to high turbulent gas surface densities maintained by cold streams at high-z
(Bournaud et al. 2011).

The availability of gas, or lack thereof, is thus an essential
ingredient for feeding the black hole, and mechanisms affecting the
gas are also believed to influence the AGN (e.g. Sabater, Best \&
Heckman 2015). Given that the
gas content of galaxies is especially sensitive to environmental effects,
AGN studies as a function of enviroment are of interest, though
they often find contrasting results
(Miller et al. 2003, Kauffmann et al. 2004, Martini et al. 2006,
Popesso \& Biviano 2006, von
der Linden et al. 2010, Marziani et al. 2016).

GASP can investigate the link between the gas availability, 
gas physical conditions, and AGN activity. The
IFU data permit an investigation of the galaxy
central regions of all galaxies, a detailed analysis of the gas ionization
source (thanks to the large number of emission lines included in the
spectra) and the detection of eventual AGN-driven large-scale outflows.
Future GASP papers will present the occurrence of AGN among
ram-pressure stripped galaxies (e.g. Poggianti et al. 2017b submitted).
As an example, the galaxy presented in this paper hosts an AGN (\S 7.2).

\subsection{Galaxy evolution without and before stripping}
The GASP control sample consists of disk galaxies with a range of
galaxy masses and in different environments, with no sign of
disturbance/debris. At all effects, they can be considered a
sample of ``normal galaxies''.
Moreover, the GASP stripping candidates undergoing gas-only removal
processes have their stellar component undisturbed, retaining the
memory of the galaxy history before stripping. Thus, from the
stellar component of the MUSE datacube with our
spectrophotometric code we can recover the past galaxy history at times before the
stripping occurred (see \S 6.2).

Thus, GASP can be used to derive the spatially
resolved stellar and metallicity history in absence, or prior to,
galaxy removal, as well as the ongoing star formation activity and
ionized gas properties in normal galaxies.

Compared to other larger IFU surveys (S\'anchez et al. 2012, Allen et
al. 2015, Bundy et al. 2015), GASP has the disadvantage of the
smaller number of galaxies, but the advantage of covering many
galactic effective radii. The outer regions of galaxies hold a unique set of clues
about the way in which galaxies are assembled (Ferguson et al. 2016).  
With GASP it is possible to peer into galaxy outskirts, to study
the stellar, gas and dust
content out to large radii in galaxies, enabling to compare 
the star formation history and metallicity gradients 
with simulations of disk galaxy formation (Mayer et al. 2008,
  Vogelsberger et al. 2014, Kauffmann et al. 2016, Christensen et al. 2016).
GASP observations are suitable to investigate
how star formation occurs at
low gas densities and low metallicities,
and may hold clues about stellar
migration and satellite accretion.

\section{Survey strategy}

\subsection{Parent surveys: WINGS, OMEGAWINGS and PM2GC} 
The GASP program is based on three surveys that, together, cover the
whole range of environmental conditions at low redshift: WINGS,
OMEGAWINGS and PM2GC.

WINGS is a multiwavelength survey of 76 clusters of galaxies at
z=0.04-0.07 in both the north and the south emisphere
(Fasano et al. 2006). The clusters were selected on the
basis of their X-ray luminosity (Ebeling et al. 1996, 1998, 2000) and
cover a wide range in halo mass ($10^{13.6}-10^{15.2} \, M_{\odot}$), with
velocity dispersions $\sigma=$ 500-1300 $\rm km \, s^{-1}$ and X-ray
luminosities $\rm L_X = 10^{43.3-45} \, \rm erg \, s^{-1}$.  The
original WINGS dataset comprises deep B and V photometry with a
$34^{\prime} \times 34^{\prime}$ field-of-view with the WFC@INT and
the WFC@2.2mMPG/ESO (Varela et al. 2009), spectroscopic follow-ups
with 2dF@AAT and WYFFOS@WHT (Cava et al. 2009), J and K imaging
with WFC@UKIRT (Valentinuzzi et al. 2009) and U-band imaging (Omizzolo et
al. 2014). The WINGS database is presented in Moretti et al. (2014)
and is all publicly available through the Virtual Observatory.

OMEGAWINGS is a recent extension of the WINGS project that has
quadrupled the area covered in each cluster (1 square degree). 
B and V deep imaging with OmegaCAM@VST was secured 
for 46 WINGS clusters (Gullieuszik et al. 2015), a $u$-band
program is ongoing with the same instrument (D'Onofrio et al. in
prep.) and an AAOmega@AAT spectroscopic campaign yielded 
18.000 new redshifts (Moretti et al. 2017) together with
stellar population properties and star formation rates  that have been
used in Paccagnella et al. (2016).

As a comparison field sample we use the Padova Millennium Galaxy and
Group Catalogue (PM2GC, Calvi et al. 2011), which is drawn from the 
Millennium Galaxy Catalogue (MGC, Liske et al. 2003). The MGC data
consist of deep B-band imaging with WFC@INT over a
38 $deg^2$ equatorial area and a highly complete spectroscopic
follow-up (96\%  at B=20, Driver et al. 2005). 
The PM2GC galaxy sample is thus representative of the general field
and as such contains galaxy groups (176 with at least three members at
z=0.04-0.1), binary systems and single galaxies, as identified with a
Friends-of-friends algorithm by Calvi et al. (2011), covering a
braod range in halo masses (Paccagnella et al. in prep.).



\subsection{Selection of GASP targets}

GASP is planning to observe 114 galaxies, of which 94 are primary targets and 20
compose a control sample.

\subsubsection{Primary targets: the Poggianti et al. (2016) atlas}
GASP primary targets were taken from the atlas of Poggianti et
al. (2016, hereafter P16), who provided a large sample of galaxies 
whose optical morphologies are suggestive of gas-only removal mechanisms.
These authors visually inspected B-band WINGS, OMEGAWINGS and PM2GC images searching for 
galaxies with (a) debris trails, tails or surrounding debris
located on one side of the galaxy, (b) asymmetric/disturbed
morphologies suggestive of unilateral external forces or (c) a
distribution of star-forming regions and knots suggestive of triggered
star formation on one side of the galaxy. 
Galaxies whose morphological disturbance was
clearly induced by mergers or tidal interactions were
deliberately excluded. The selection was based only on the
  images, therefore a subset of the candidates did not have a 
  known spectroscopic redshift.
For the PM2GC, only galaxies with a spectroscopic redshift 
in the range of WINGS clusters (z=0.04-0.07) were considered.

P16 classified candidates according to the strength
of the optical stripping signatures: JClass 5 and 4 are those with the
strongest evidence and are the most secure candidates, including
classical ``jellyfish galaxies'' with tentacles of stripped
material; JClass 3 are probable cases of stripping, and JClass 2 and 1
are the weakest, tentative candidates.

After inspecting a total area of about 53 $deg^2$ in clusters
(WINGS+OMEGSAWINGS) and 38
$deg^2$ in the field (PM2GC),  the total P16 sample consists of 344
stripping candidates in clusters
(WINGS+OMEGAWINGS) and 75 candidates in the field (PM2GC),
finding apparently convincing candidates for gas-only removal
mechanisms also in groups down to low halo masses. 
While for cluster galaxies the principal culprit is commonly assumed
to be ram pressure stripping, this is believed to be too inefficient
in groups and low-mass haloes, where other mechanisms, such as
undetected minor or major mergers or tidal interactions, might give rise to similar
optical features. The integral field spectroscopy obtained by GASP is the optimal method
to identify the physical process at work, because it probes both gas
and stars and can discriminate processes affecting only the gas, such
as ram pressure, from those affecting gas and stars, such as tidal
effects and mergers.

The P16 candidates are all disky galaxies with stellar masses in the range $\sim
10^{9}-10^{11.5} \, M_{\odot}$ and are mostly star-forming with a star
formation rate that is enhanced on average by a factor of 2 compared
to non-candidates of the same mass.

GASP will observe as primary targets 64
cluster and 30
field stripping candidates taken from the P16 atlas. 
They are selected according to the following 
criteria: (a) be observable from Paranal (DEC$<+15$); (b) include all the 
JClass=5 objects; (c) include galaxies of each JClass, from 5 to 1, in 
similar numbers; (d) for WINGS+OMEGAWINGS, give preference to 
spectroscopically confirmed cluster members, rejecting known 
non-members; (e) cover the widest possible galaxy mass range. 

Spatially resolved studies of gas in cluster galaxies have
shown that the stripping signatures visible in the optical images
are just the tip of the iceberg: the stripping is much more evident
from the ionized gas observations than in the
optical (Merluzzi et al. 2013, Fumagalli et al. 2014, Kenney et
al. 2015). For this reason we chose to include in our study galaxies
with a wide range of degree of evidence for stripping (of all
Jclasses plus a control sample), to obtain a complete view of gas removal
phenomena.

\begin{figure*}
\centerline{\includegraphics[scale=0.55]{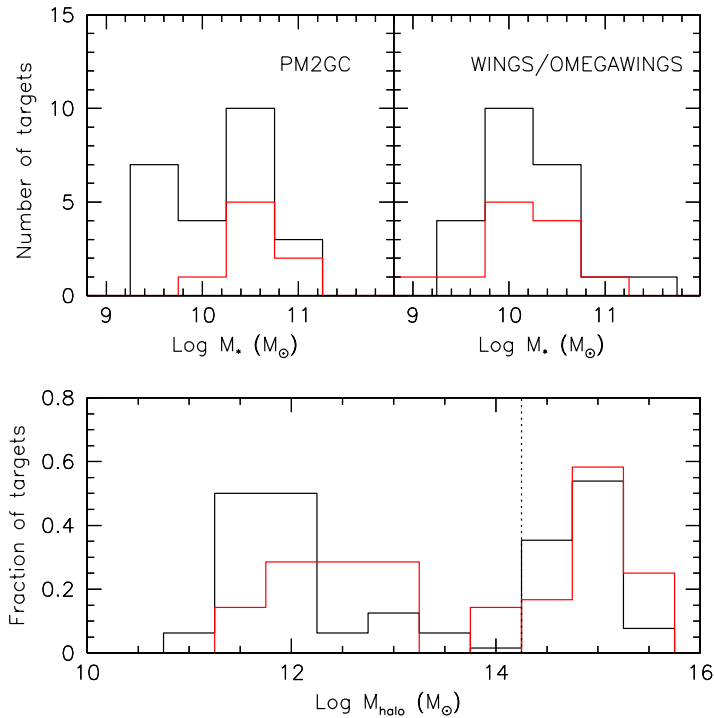}}
\vspace{-3cm}
\caption{Top. Galaxy stellar mass distributions of cluster (right, WINGS/OMEGAWINGS) and
  group/field targets (left, PM2GC) for the primary sample (solid black
  histogram) and the control sample (dashed green histograms).
Bottom. Host halo mass distributions of WINGS/OMEGAWINGS (to the right
of the dotted horizontal line) and PM2GC (to the left of the line)
targets. All these distributions are subject to change (see text).}
\end{figure*}

Figure~1 presents the galaxy stellar mass and host
halo mass distributions of GASP primary and control targets, that are
similar to those of their parent samples in Poggianti et al. (2016). 
At this stage these distributions are preliminary and incomplete, but
already allow the reader to gauge the range of masses covered. They are
preliminary because the final sample that will be observed may not be
exactly the same we envisage at the moment, due to observability,
scheduling etc. in service mode; they are incomplete because
for now we have galaxy mass estimates only for the subset of targets
with WINGS/OMEGAWINGS/PM2GC spectroscopy, while MUSE will provide
masses for each galaxy.
Final properties of the sample, together with the complete list of
targets, will be published in subsequent paper of this series once the
observations are finalized.


\subsubsection{The control sample}
Galaxies in the control sample are galaxies in clusters (12) and in the
field (8) with {\it no} optical sign of stripping, i.e. no signs of debris or
unilaterally disturbed morphologies in the optical images.
This sample will allow to contrast the properties of stripping
candidates with those of galaxies that show no optical evidence of
gas removal. Finding signs of gas stripping from the IFU data
even in galaxies of the control sample would reveal that the stripping phenomenon is more
widespread than it is estimated from the optical images. Moreover, the control sample
represents a valuable dataset of  ``normal'' disk galaxies that
allows a spatially resolved study out to several galaxy effective radii.

Control sample galaxies were selected from WINGS, OMEGAWINGS and PM2GC 
visually inspecting the same B-band images used for the primary targets and
were chosen according to the following criteria: (a)  be at DEC$<+15$
(b) for WINGS+OMEGAWINGS, be spectroscopically confirmed cluster
members; (c) have a stellar mass estimate from the spectral fitting 
(Paccagnella et al. 2016 for WINGS+OMEGAWINGS, Calvi et al. 2011 for
PM2GC) and cover a galaxy mass range as similar as possible to that 
of the primary targets (Fig.~1); (d) be
spirals spanning the same morphological range of the primary targets
(Sb to Sd), with the addition of a few lenticulars and early spirals
for comparison. For both WINGS/OMEGAWINGS and PM2GC, the morphologies are 
derived with the MORPHOT automatic classification tool (Fasano et
al. 2012, Calvi et al. 2012); (e)
include both star-forming (emission-line), post-starburst (k+a/a+k)
and passive (k) spectral types according to the definition in Fritz et al. (2014).

\section{Observations and data reduction}

Observations are currently undergoing and are carried out in service
mode 
using the MUSE spectrograph located at the Nasmyth
B focus of the {\it Yepun} (Unit Telescope 4)  VLT in Paranal. The constraints demanded for the observations are: clear conditions,
moon illumination $< 30\%$, moon distance $> 30$ degrees , and image
quality $< 0.9 \rm \, arcsec$, corresponding to $<1 \rm \,arcsec$
seeing at zenith. 

MUSE (Bacon et al. 2010) is composed of 24 IFU modules, equipped with
24$\times$4k$\times$4k CCDs. We use the MUSE wide-field mode 
with natural seeing that covers approximately a 1'$\times$1' field-of-view
with 0.2''$\times$0.2'' pixels.
The spectral range between 4800 \AA $\,$ and 9300 \AA $\,$ is sampled
with a resolution FWHM$\sim 2.6$ \AA $\,$ (R=1770 at 4800 \AA $\,$ and
=3590 at 9300 \AA $\,$) and a sampling of 1.25 \AA/pixel.
Thus, each datacube yields approximately 90.000 spectra.

\subsection{Observing strategy}
The majority of GASP galaxies are observed with 4 exposures of
675 sec each, each rotated by 90 degree and slightly offsetted
with respect to the previous one, to minimise the cosmetics. 
The minimum time on target is
therefore 2700 sec per galaxy. 
Some targets, however, show long tails in the optical images and
require two, or even three, offsetted pointings to cover the galaxy
body and the lenght of the tails. Each of these pointings is covered
with 2700 sec, split into 4 exposures, as above.

The great majority of pointings have a significant fraction of sky coverage, while
for a few galaxies it is necessary to do a 120 sec sky offset after each 
675 sec exposure, because the galaxy fills the MUSE FOV. 

Standard calibration frames are taken for each observation according
to the ESO MUSE Calibration Plan. In short, at least one spectrophotometric
standard star was observed each night for flux and telluric correction
purpose. An internal illumination correction flat is also taken near
the beginning or the end of the observations to minimise flatfielding
issues due to ambient temperature changes. Daytime calibrations such
as arcs, biases, darks, and flats (both internal and sky) are also
taken. Static calibrations such as astrometry, line-spread-functions 
etc come as default in the pipeline supplied by ESO. 

\subsection{Data reduction}



The data are reduced with the most recent version of the MUSE pipeline
at any time (Bacon et al. 2010, https://www.eso.org/sci/software/pipelines/muse/). This was version 1.2 for the first data taken, and is
v.1.6 at the time of writing. The procedures and philosophy of the
data-reduction follow closely those set out in the ESO Pipeline
Manual. To speed up and automate the process, raw data are organised
and prepared with custom scripts, then fed to ESOREX recipes
v.3.12. For most observations, the pipeline can be run in a semi
automated fashion, since the observations are mostly identical in the
execution and calibration. Briefly, the pipeline is run with mainly
default parameters. The data and the standard star frames are
flat-fielded, wavelength calibrated and corrected for differential
atmospheric refraction. Typical wavelength calibration
has $\approx 0.025$ \AA$\,$ RMS in the fit, and the mean resolution $R$
measured from the arcs is about 3000.

As explained above, most of the exposures have sufficient sky coverage
within the MUSE field-of-view, leaving $>50\%$ area for sky
measurements, and the sky is modeled directly from the individual
frames using the $20\%$ pixels with the lowest counts, thus there is
no risk of accidentally subtracting any faint diffuse H$\alpha$ within
the FOV. For spatially extended galaxies the offset sky exposures of 
120s allow the sky to be modelled adequately. 

The standard star observation closest in time to the science
observations is used for the flux calibration. After flux calibration
and telluric correction, the final flux-calibrated datacube is
generated by lining up the individual frames using sources in the
white-light images to calculate the (small) offsets. Galaxies with
multiple pointings use sources in the overlaps for alignment.
In a few cases we found no sources in the overlap: we therefore
computed the offsets using custom scripts and OMEGAWINGS images 
as reference.


MUSE spectra in the red, where strong skylines dominate, are known to
have 
residuals of sky subtraction in
the current pipeline implementation. As the most interesting
absorption and emission lines for our galaxies lie blueward of
7200 \AA, this issue does not pose a problem. When the red part of the
spectrum is needed for analysis, we perform a further cleaning of the
spectrum redward of 7200 \AA$\,$ using ZAP (Soto et al 2016) which provides
satisfactory results. However experimentation of ZAP shows that it has
to be used with caution, especially if there may be faint extended
$\rm H\alpha$ emission in the “sky” spaxels which ZAP can aggressively
“clean”. 


\section{Current status and data products release policy}

At the time of writing, in January 2017, 
55 out of the 114 targets have been observed. All data taken have been
reduced. Based on the current
rate of execution, we project a completion by the end of 2018.

GASP is an ESO Large Program committed to release its products into
the ESO Science Archive. These products will include input target
catalogs, fully reduced and calibrated MUSE datacubes, catalogs with
redshifts and emission-line fluxes for galaxy spaxels and catalogs of
outputs of our spectrophotometric model SINOPSIS (Fritz et
al. submitted, Paper III) with stellar masses, 
stellar ages and star formation histories. 


\section{Data analysis}

This section describes the procedure we use to analyze all galaxies of the survey. 
The chart in Fig.~2 presents the work-flow
described below. In the following, we first describe the spectral
analysis spaxel by spaxel, then we consider the integrated spectra of individual star-forming
knots (\S6.4) and of the galaxy main body (\S6.5). Extraplanar knots of star formation
turn out to be common in GASP galaxies and therefore are an important
aspect of our analysis.

\begin{figure*}
\centerline{\includegraphics[scale=0.35]{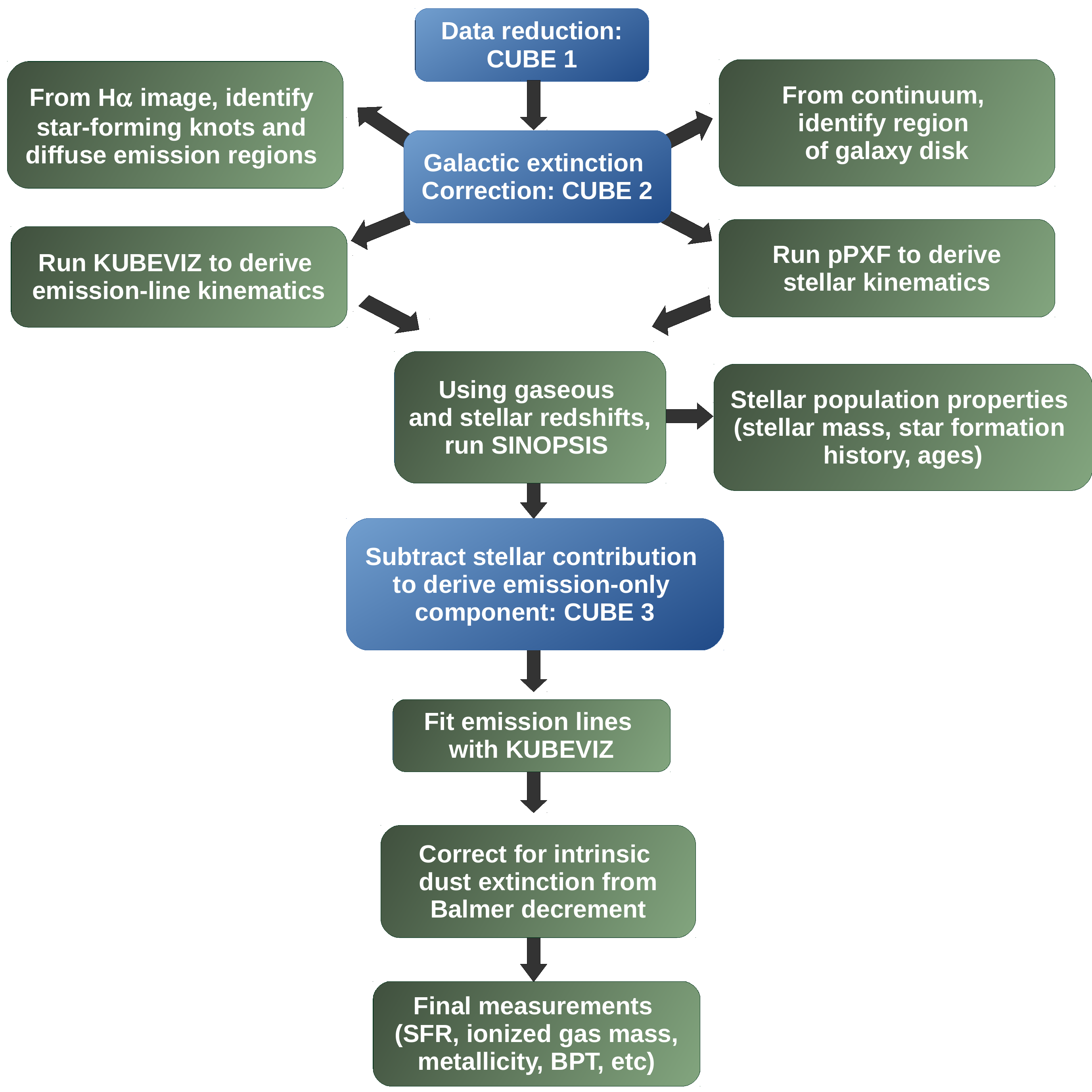}}
\caption{Schematic chart of the GASP analysis work-flow (see text for details).}
\end{figure*}

First of all, the reduced datacube is corrected for extinction due to our
own Galaxy, using the extinction value estimated at the galaxy position
(Schlafly \& Finkbeiner 2011) and assuming the extinction law from Cardelli et al. (1989).
The corrected datacube (CUBE 2 in Fig.~2) is used in all the subsequent analysis.

\subsection{Gas and stellar kinematics}

To analyze the main emission lines in the spectrum we use the
IDL publicly available software KUBEVIZ (Fossati et al. 2016), written 
by Matteo Fossati and David Wilman.
Starting from an initial redshift, KUBEVIZ uses the MPFit (Markwardt 2009)
package to fit gaussian line profiles, yielding gaseous velocities (with
respect to given redshift), velocity dispersions and total line fluxes. The list of
emission lines fitted by KUBEVIZ is given in Table~1, together with
the characteristic surface brightness $3\sigma$  limit for each line.

\begin{table}
\centering
\caption{Emission lines fitted with KUBEVIZ for this study.} 
\begin{tabular}{lcc}
\hline
Line & $\lambda$  &  $\mu_{lim}$ \\
        & (\AA) (air) &  $\rm erg \, s^{-1} \, cm^{-2} \, arcsec^{-2}$\\
\hline
$\rm H\beta$ & 4861.33  &  0.3 -- 1.6 $\times 10^{-17}$ \\
$\rm [OIII]$ &  4958.91   & 0.3 -- 1.2 $\times 10^{-17}$  \\
$\rm [OIII]$ &  5006.84   & 0.3 -- 1.2 $\times 10^{-17}$   \\
$\rm [OI]$ &    6300.30 &  0.2 -- 0.8 $\times 10^{-17}$  \\
$\rm [OI]$ &    6363.78  & 0.3 -- 0.9 $\times 10^{-17}$       \\
$\rm [NII]$ &   6548.05    & 0.2  -- 1.0 $\times 10^{-17}$      \\
$\rm H\alpha$ &   6562.82    &  0.2 --  1.1 $\times 10^{-17}$     \\
$\rm [NII]$ &   6583.45   &  0.2 --  1.0 $\times 10^{-17}$       \\
$\rm [SII]$ &    6716.44   & 0.2  --   0.8 $\times 10^{-17}$      \\
$\rm [SII]$ &   6730.81  &   0.2 --   0.8 $\times 10^{-17}$      \\
\hline
\end{tabular}
\tablecomments{For each line we list the wavelength (air) 
and the characteristic surface brightness limit (3$\sigma$) in each spaxel 
after Galactic extinction correction. The two values refer to the 
limits from the datacube used for the analysis (average filtered in spatial direction 
with a $5\times5$ pixel kernel, see text) and from the original 
datacube ($1\times1$).}
\end{table}

KUBEVIZ uses ``linesets'',
defined as groups of lines that are fitted simultaneously. Each
lineset (e.g. $\rm H\alpha$  and [NII]6548, 6583) is considered a combination
of 1D Gaussian functions keeping the velocity separation of the lines
fixed according to the line wavelengths. KUBEVIZ imposes a prior on
the velocity and instrinsic line width of each lineset, which is fixed
to that obtained by the fit of the $\rm H\alpha$ and [NII] lines.
Moreover, the flux ratios of
the two [NII] and [OIII] lines are kept constant in the fit assuming
the ratios given in Storey \& Zeippen (2000).

Before carrying out the fits, the datacube is average filtered in the
spatial direction with a 5$\times$5 pixel kernel, corresponding to
1 arcsec=0.7-1.3 kpc depending on the galaxy redshift.
Moreover, as recommended by Fossati et al. (2016), the errors on the
line fluxes are scaled to achieve a reduced ${\chi}^2=1$.
 KUBEVIZ can attempt a single or a double component fit. A single fit
 was run for each galaxy, while another KUBEVIZ run with a double
 component was necessary in some cases, as described in Paper II.
The continuum is calculated between 80 and 200 \AA $\,$ redwards and
bluewards of each line, omitting regions with other emission lines and
using only values between the 40th and 60th percentiles.

Maps of $\rm H\alpha$ intensity, velocity and velocity dispersion are
created at this stage using the KUBEVIZ output. The original datacube is visually
and carefully inspected, and contrasted with these maps, for a)
finding fore- and back-ground sources that are superimposed on the
galaxy of interest along the line of sight. A mask of these sources is
created in order to remove the contaminated regions from the stellar
analysis described below and, b) checking the output of KUBEVIZ and
ensuring all lines are correctly identified.


To extract the stellar kinematics from the spectrum, we use the Penalized
Pixel-Fitting (pPXF) code (Cappellari \& Emsellem, 2004),
fitting the observed spectra with the stellar population templates by Vazdekis
et al. (2010).
We used SSP of 6 different metallicities (from $[M/H]=-1.71$ to $[M/H]=0.22$)
and 26 ages (from 1 to 17.78 Gyr) calculated with 
the Girardi et al. (2000) isochrones. 
Given the poor resolution of the theoretical stellar libraries in the red part of the spectra, 
and the strong contamination from the sky lines in the observed
spectra redwards of the $\rm H\alpha$ region, 
we run the code on the spectra cut at $\sim 7000$ \AA $\,$ restframe. 
We first accurately masked spurious sources (stars, 
background galaxies) in the galaxy proximity, that could enter one of the  
derived bins.  
After having degraded the spectral library resolution to our MUSE
resolution, we first find spaxels belonging to the galaxy, here defined
as those with a signal median-averaged over all wavelengths 
larger than $3 \times 10^{-20} \, \rm erg s^{-1} cm^{-2} {\AA}^{-1}$ per
  spaxel. 
Then we performed the fit of spatially binned spectra based
on signal-to-noise (S/N=10, for most galaxies), as described
in Cappellari \& Copin (2003), with the Weighted Voronoi Tessellation 
modification proposed by Diehl \& Statler (2006).
This approach allows to perform a better tessellation in case of non
Poissonian noise, not optimally weighted pixels and when the 
Centroidal algorithm introduces significant gradients in the S/N.

We derived maps of the rotational velocity, the velocity dispersion and the two h3 and
h4 moments using an additive Legendre polynomial fit of the 12th order
to correct the template continuum shape during the fit.
This allows us to derive for each Voronoi bin a redshift estimate, that was then
used as input for the stellar population analysis.

\subsection{Emission line fluxes and stellar properties}
To obtain the measurements of total emission line fluxes, corrected
for underlying stellar absorption, and for deriving spatially
resolved stellar population properties we run our 
spectrophotometric model SINOPSIS (Paper III).
This code searches the combination of single stellar population
(SSPs) spectra that best fits the equivalent widths of the main lines in absorption
and in emission and the continuum at various wavelengths,
minimizing the ${\chi}^2=1$ using an Adaptive Simulated Annealing
algorithm (Fritz et al. 2011, 2007).
The star formation history is let free with no analytic priors.

The code, which has been employed to derive star formation histories and
rates, stellar masses and other stellar properties of various surveys
(Dressler et al. 2009, Fritz et al. 2011, Vulcani et al. 2015, Guglielmo et al. 2015,
Paccagnella et al. 2016, Cheung et al. 2016)
has been substantially
updated and modified for the purposes of GASP. 
The GASP version of SINOPSIS is fully described in Paper III,
here we only describe the main improvements
with respect to Fritz et al. (2011).
The code now uses the latest SSPs model from Charlot \& Bruzual (in
prep.) with a higher spectral and age resolution. They use a Chabrier
(2003) IMF with stellar masses in the 0.1–-100 M$_\odot$ limits, and
they cover metallicity values from $Z=0.0001$ to $Z=0.04$. These
models use the latest evolutionary tracks from Bressan et al. (2012) and
stellar atmosphere emission from a compilation of different authors,
depending on the wavelength range, and on the stellar
luminosity and effective temperature. In addition, nebular emission
has been added for the youngest (i.e.age $< 2\times 10^7$ years) SSP,
by ingesting the original models into CLOUDY (Ferland et al. 2013). In
this way, the SSP spectra we use display also the most common and most
intense emission lines (e.g. Hydrogen, Oxygen, Nitrogen). Moreover, the code
has been improved and optimised to deal efficiently with datacubes
such as the products from MUSE. It is now possible to read in the
observed spectra directly from the cube fits file, while the redshifts
for each spaxel are taken from 2D redshift masks. 


SINOPSIS requires the spectrum redshift as input, thus the redshift at
each location of the datacube was taken from pPXF (stellar) and
KUBEVIZ (gaseous) as described above.
As the stellar and gas components might be kinematically decoupled, 
the observed wavelength of a given line in emission (gas) could differ 
from that of the same line in absorption (stellar photosphere). 
This might result in an erroneous measurement of the line, 
depending on which redshift is adopted, introducing issues 
most of all for the H$\beta$ line, where the emission and 
absorption components can be separately identified. 
Hence, we have introduced a further option in SINOPSIS to allow the
use of the gas redshift, 
when available, to detect and measure the equivalent width of 
emission lines, while the stellar redshift is used to fit the continuum and measure absorption lines.
The code is then run on the 5$\times$5 pixel average
filtered observed cube.

SINOPSIS produces a best fit model cube (stellar plus gaseous emission) and a stellar-only
model cube (containing only the stellar component). The latter is
subtracted from the observed datacube to provide an emission-only
cube (CUBE 3 in Fig.~1) that is used for all the following
analysis. KUBEVIZ is run a second time ($\rm KUBEVIZ_{run2}$) on this emission-only cube to
estimate the emission line fluxes corrected for stellar absorption.

In addition, SINOPSIS gives spatially resolved 
estimates of the
stellar population properties, and maps are produced for:
a) stellar masses; b) average star formation rate and total mass formed in 4 age bins:
young(ongoing SF)=$< 2 \times 10^7$ yr;
recent=$2 \times 10^7 < 5.7 \times 10^8$ yr;
intermediate age= $5.7 \times 10^8 < 5.7 \times 10^9$ yr;
old=$> 5.7 \times 10^9$ yr; c) luminosity-weighted stellar ages.


\subsection{Derived quantities: dust extinction, gas metallicity,
  diagnostic diagrams, star formation rates and gas masses}

The emission-line, absorption-corrected fluxes measured by $\rm
KUBEVIZ_{run2}$ are
then corrected for extinction by dust internal to the galaxy. 
The correction is derived from the Balmer decrement at each spatial
element location assuming an intrinsic $\rm H\alpha$/$\rm H\beta$ ratio equal
to 2.86 and adopting the Cardelli et al. (1989) extinction law.
This yields dust- and absorption-corrected emission line fluxes and a
map of the dust extinction $A_V$.
These fluxes are then used to derive
all the quantities discussed below.


%
%
The gas metallicity and $q$ ionization parameter
are calculated at each
spatial location using the {\it pyqz} code (Dopita et al. 2013) \footnote{\url{http://fpavogt.github.io/pyqz}} 
v0.8.2.  The $q$ parameter is
  defined by Dopita et al. (2013) as the ratio between the number of
  ionizing photons per unit area per second and the gas particle number
density \footnote{ It is $U=q/c$, where $U$ is the classic definition of the
ionization parameter and $c$ is the speed of light.}.
The code
interpolates a finite set of diagnostic line ratio grids computed with
the MAPPINGS code to compute $\log(Q)$ and $12+\log(O/H)$.  The
MAPPINGS V grids in pyqz v0.8.2 cover a limited range in
abundances ($8.11 \leq 12+log(O/H) \leq 8.985$); we therefore used a
modified version of the code to implement MAPPINGS IV grid tested in
the range $7.39 \leq 12+log(O/H) \leq 9.39$ (F. Vogt,
priv. communication).  We used the calibration based on the strong
emission lines, namely [NII]6583/[SII]6716,6731 vs [OIII]5007/[SII]6716,6731 \footnote{[SII] is
the sum of the fluxes of the two [SII] lines.}.  
Using only one diagnostic could in principle lead to systematic effects both 
on the ionization parameter and on the metallicity values, as
shown in Dopita et al. (2013). However, conclusions based on differential analyses of metallicity 
variations/gradients within a galaxy are not affected by this problem.

The line fluxes are also used to create
line-ratio diagnostic diagrams (Baldwin et al. 1981, BPT) to investigate the cause of the gas
ionization and distinguish regions photoionized by hot stars from regions ionized
by shocks, LINERs and AGN. Only spaxels with a S/N$>3$ in all the
emission lines involved are considered.
For each galaxy we inspect and compare the conclusions based on three diagrams:
[OIII]5007/$\rm H\beta$ vs [NII]6583/$\rm H\alpha$, [OIII]5007/$\rm 
H\beta$ vs [OI]6300/$\rm H\alpha$
and [OIII]5007/$\rm H\beta$ vs [SII]6717,6731/$\rm H\alpha$. 

The SFR of each spatial element is computed from the $\rm H\alpha$
luminosity corrected for dust and stellar absorption adopting Kennicutt (1998)'s relation: 
\begin{equation}
SFR = 4.6 \times  10^{-42} L_{H\alpha} 
\end{equation}
where SFR is in solar masses per year and the $\rm H\alpha$ luminosity
is in erg per second, for a Chabrier IMF. 

The total SFR is computed from the sum of the
dust-corrected $\rm
H\alpha$ fluxes in each spaxel, after removing hot pixels and adopting
a S/N cut (between 3 and 7).
The same method is used to compute the
SFR within the ``galaxy body'' (i.e. the stellar outer isophotes
described in sec.~6.5). The latter
can also be computed with SINOPSIS from the integrated spectrum,
but only without removing the AGN contribution.

The  $\rm H\alpha$ luminosity can be employed to estimate the
mass of ionized gas (e.g. Boselli et al. 2016).
From Osterbrock \& Ferland (2006, eqn. 13.7, pag. 344):
\begin{equation}
 L_{H\alpha}  = n_e n_p V f \alpha_{H\alpha} h \nu_{H\alpha}
\end{equation}
where V is the volume, f the filling factor, $n_e$ and $n_p$ are the
density of electrons and protons, $\rm \alpha_{H\alpha}$ is the effective
$\rm H\alpha$ recombination coefficient (
$1.17 \times 10^{-13} cm^{3} \, s^{-1}$) and $\rm h \nu_{H\alpha}$ is the energy
of the $\rm H\alpha$ photon (
$0.3028 \times 10^{-11} erg$) for a case B recombination, n=10000 $\rm cm^{-3}$,
and T=10000 K. It is commonly assumed that
$n_e$=$n_p$=$n$ (the gas is fully ionized, e.g. Boselli et
  al. 2016, Fossati et al. 2016).

The mass of ionized gas is the number of hydrogen atoms (=number of
protons) times the mass of the hydrogen atom $m_H =1.6735 \times
10^{-24} gr$. The number of protons is equal to the density of protons
times the volume times the filling factor $N_{protons}= n_p V f$.
Using eqn.(2) above we compute the ionized gas mass as:

\begin{equation}
  M_{gas} = N_{protons} \times m_p =
\frac{L_{H\alpha} \times m_p}{n \alpha_{H\alpha} h \nu_{H\alpha}}
\end{equation}

The density $n$ can be derived from the ratio of the [SII]6716 and
[SII]6732 lines. We use the calibration from Proxauf et al. (2014)
for T=10000 K,
obtained with modern atomic data using CLOUDY,
which is valid for the interval $R=[SII]6716/[SII]6732 = 0.4 - 1.435$.



\subsection{Star-forming knots}
The majority of GASP galaxies present bright star-forming knots in the
gaseous tails and/or in the galaxy disk. 


The location and radius of these knots are found through a purposely
devised shell script 
that includes IRAF and FORTRAN calls.
In the first step, the centers of knot candidates are identified as local
minima onto the laplace+median filtered $\rm H\alpha$ image derived
from the MUSE datacube (IRAF-laplace and
IRAF-median tools). 
A "robustness index" is then associated to each local minimum, based
on the gradient concordance towards the knot center for pixels around
the minimum. The final catalog of knot positions includes the local minima whose "robustness index" exceeds a given value.
In the second step the knot radii are estimated from the original $\rm H\alpha$
image and their photometry is performed by a purposely devised FORTRAN code. 
In particular, the blob's radii are estimated through
a recursive (outwards) analysis of three at a time, consecutive
circular shells (thickness: 1 pixel) around each knot center. The
iteration stops and the knot's radius is recorded when at least one
among the following circumstances occurs for the current outermost
shell:
(a) the counts of at least one pixel exceed those of the central pixel; (b) the fraction of pixels with counts greater than the average counts of the preceding shell ('bad' pixels) exceeds a given maximum value (usually 1/3); (c) the average counts of ‘good’ (not 'bad') pixels is lower than a given threshold value, previously set for the diffuse emission; (d) the image edges are reached by at least one pixel.

The knot radii provided in this way are used to obtain,
for each knot, the following photometric quantities: (1) total counts
inside the circle defining the knot, both including and excluding the
counts below the threshold previously set for the diffuse emission
(counts of pixels belonging to different knots are equally shared
among them); (2) total counts from integration of the growth curve. 
The average shell counts tracing the growth curve are obtained using 
only the ‘good’ pixels in each shell and the growth curve is extrapolated down to the diffuse emission threshold.

Besides the centers and the above mentioned measures for
each knot, the final catalog associated to each $\rm H\alpha$ image provides a
number of useful global photometric quantities of the $\rm H\alpha$  image,
including: (a) total counts coming from the knots according to the
three measures described above; (b) total counts attributable to the
diffuse emission, both including and excluding the pixels inside the
knot's circles; (c) total counts of pixels with counts above the diffuse emission threshold, but laying outside the knot's circles.  

KUBEVIZ (run3) is then run on a mask identifying all the knots, producing
emission line fits for the integrated, emission-only spectrum of each knot.
The line fluxes are corrected for dust extinction and used to derive
for each knot diagnostic diagrams, gas metallicity estimates, star formation rate
measurements and ionized gas mass estimates with the methods
described in sec. 6.3.

\subsection{Integrated galaxy spectrum}
Finally, we integrate the spectrum over the galaxy main body to obtain
a sort of ``total galaxy body spectrum''.  
To this aim, KUBEVIZ is used to obtain a 2D image of the near-$\rm H\alpha$ continuum.
The spaxels belonging to the galaxy
main body are identified slicing this image at two different count
levels, with a surface brightness difference of $\sim$2-2.5 mag.
The outer isophote encloses essentially all the galaxy body, down to $\sim 1 \sigma$ above
the background level. Since the signal-to-noise ratio is not the same
for all galaxies, this implies that the corresponding surface brightness is different for different galaxies. 
The inner isophote contains the bright
part of the galaxy, usually 70-75\% of the total counts within the
outer isophote. It is worth noting that, due to the different morphological features
of galaxies, the inner cut has been visually chosen, varying
interactively the colormap of the galaxy image. 
This means that also the surface brightness of the inner isophote is not the same for all galaxies.

SINOPSIS is run on the galaxy integrated spectrum obtained within each
of the two isophotal contours to
derive the global stellar population properties.




\section{A textbook-case jellyfish galaxy: JO206}

The quality and characteristics of GASP data, which is very
homogeneous for all targets, is best illustrated showing
the results for one galaxy of the sample, JO206\footnote{The naming of
JO206 and all other GASP targets is taken from Poggianti et
al. (2016).} (z=0.0513, WINGS J211347.41+022834.9), which
was selected as a JClass=5 stripping candidate in the poor cluster IIZW108
(z=0.0486, Moretti et al. 2017).  JO206 is present in
both the OMEGAWINGS (Gullieuszik et al. 2015) and WINGS (Varela et
al. 2009) images, at RA=21 13 47.4 DEC= +02 28 35.5 (J2000).  An RGB
(u, B and V bands) image of the central region of this cluster is
shown in Fig.~3, with JO206 and the Brightest Cluster Galaxy (BCG)
highlighted.  No spectroscopic redshift was available 
before it was observed by GASP.

\begin{figure*}
\centerline{\includegraphics[scale=0.45]{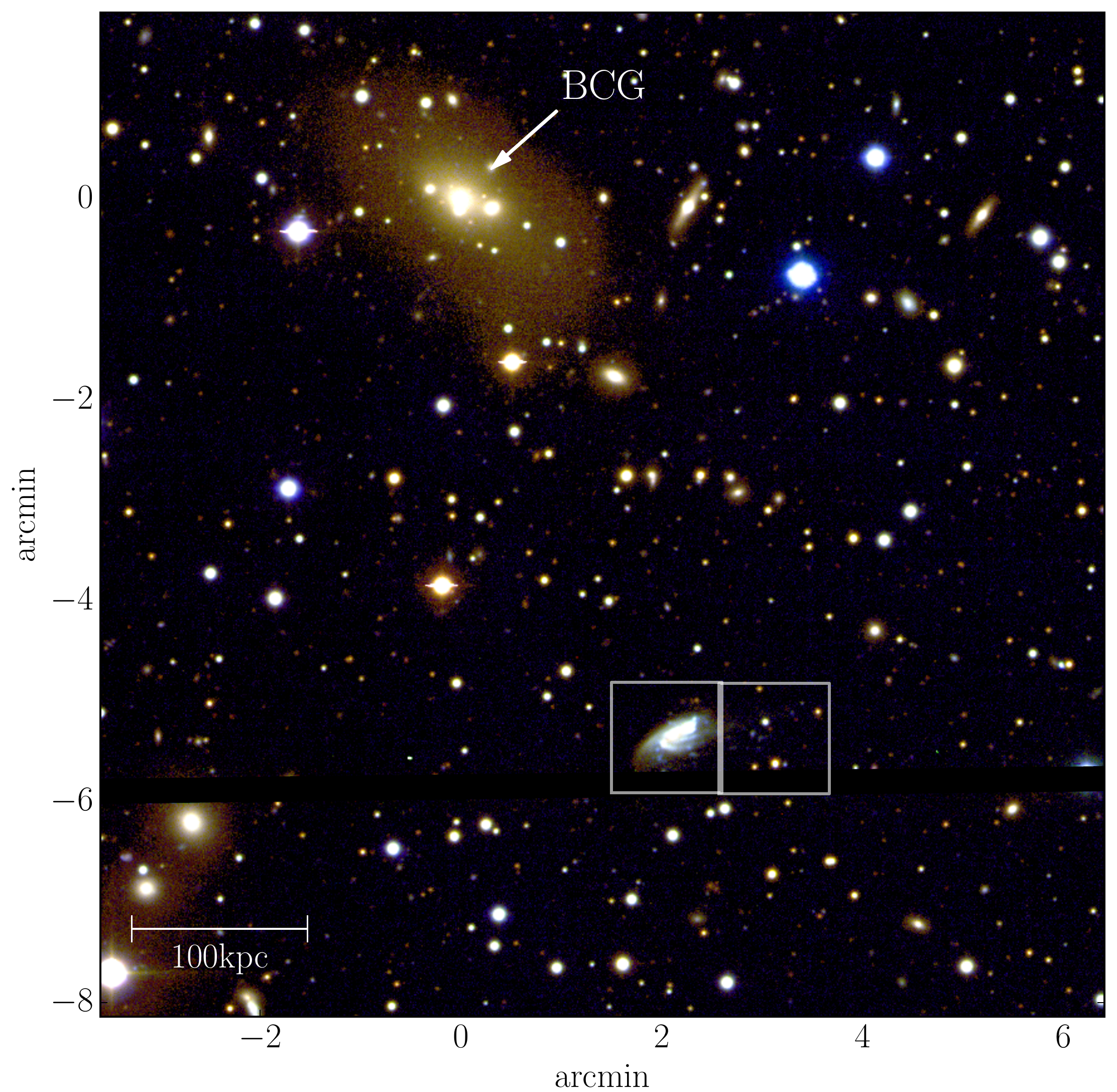}\hfill\includegraphics[scale=0.45]{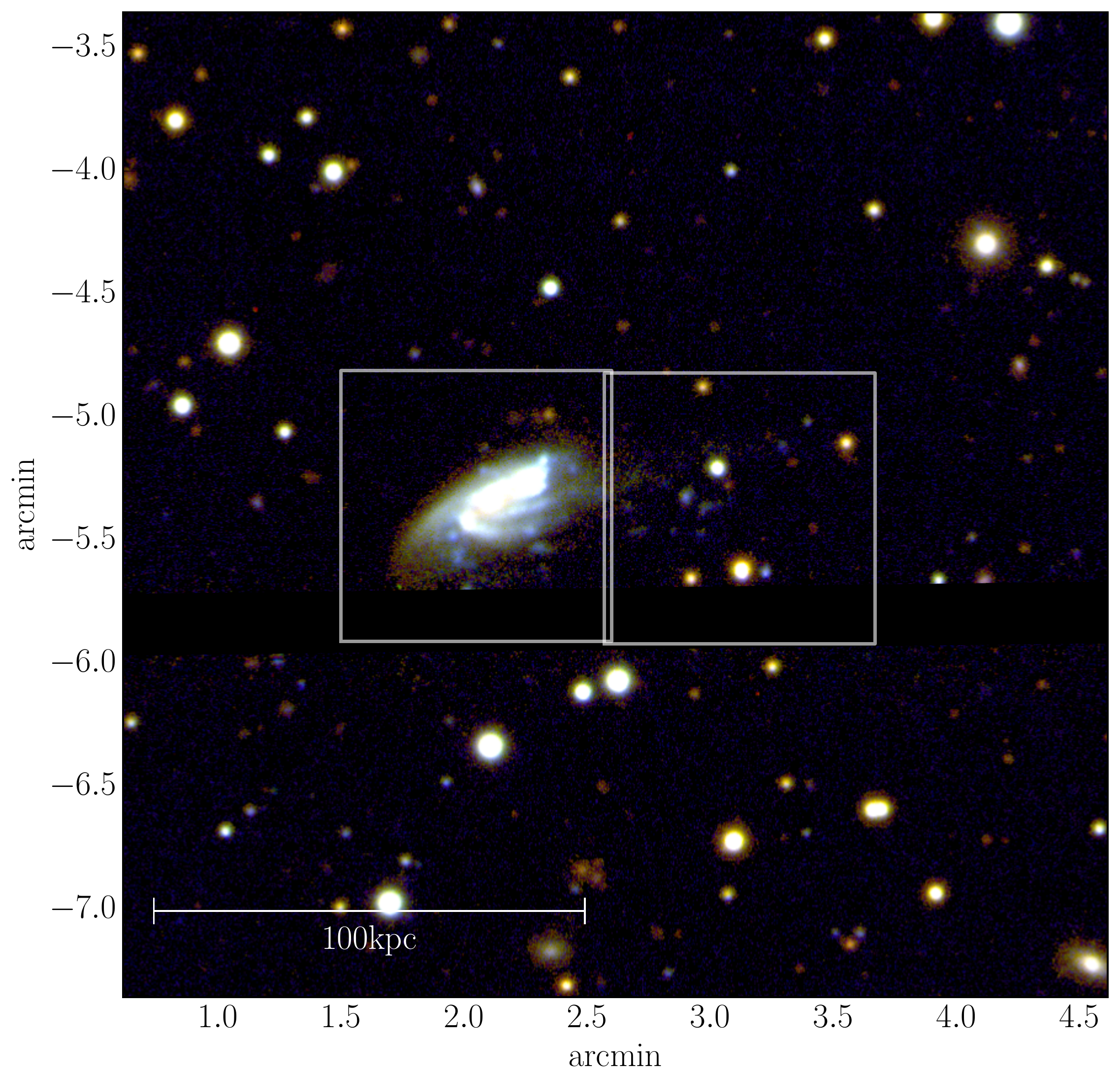}}
\caption{Left. RGB image (OMEGAWINGS u, WINGS B and V) of the central region of the cluster IIZW108, with
  the forming BCG and the two MUSE pointings for JO206
  highlighted. North is up, and east is left. Right. The same RGB image zoomed on JO206.}
\end{figure*}

JO206 was observed with two MUSE pointings ($2700$ sec  
each) on August 7 2016, with 1'' seeing during the first pointing and  
1.2 arcsec during the second pointing\footnote{In service mode at ESO,
the seeing is allowed to exceed the required value only for an OB
longer than 1 hour, as it is the case for our double
pointings. Therefore, the second pointing of JO206 has a seeing $>1
arcsec$, and this galaxy is in a sense a ``worst case'' for GASP
observing conditions.}.  
From the GASP integrated disk galaxy spectrum, defined as described in
\S 6.5, SINOPSIS 
yields a total galaxy stellar mass 8.5$\times 10^{10} M_{\odot}$
within the outer isophote, a 
total ongoing SFR=$\sim 7 M_{\odot} \, yr^{-1}$, and a luminosity weighted age 
of $\sim 1$Gyr. 

The MUSE white image (Fig.~4, 4750-9350 \AA) displays faint traces of tails with  
knots to the west of the galaxy body: they are the reason why this  
galaxy was selected as stripping candidate in the first place. 
 The extent of the stripped gas becomes much more striking 
in the MUSE $\rm H\alpha$ map (Fig.~5, left).
$\rm H\alpha$ emission is observed in the galaxy disk, in a projected
40kpc-wide extraplanar region to the south-west of the disk and in
tentacles extending 90kpc to the west, giving this galaxy the
classical jellyfish shape. Additional MUSE pointings
would be needed to investigate how far beyond the edge of the image
the tails extend. Both in the galaxy disk and in the  
stripped gas, the $\rm H\alpha$ image is characterized by regions of diffuse emission
and brighter emission knots, which will be studied individually in
\S 7.4. Moreover, in the NW region of the disk, the enhancement of $\rm H\alpha$
emission could be related to gas compression due to the ram pressure
stripping, as in Merluzzi et al. (2013).

The $\rm H\alpha$ signal-to-noise ratio (SNR) map is shown in  
Fig.~5 (right) for all spaxels with $S/N>4$.  
 The data reach a surface brightness detection limit of $V \sim
27 \rm mag \, arcsec^{-2}$ and $log \rm H\alpha \sim
-17.6 \, erg \, s^{-1} \, cm^{-2} \, arcsec^{-2}$ at the $3\sigma$ confidence level.

\subsection{Gas and stellar kinematics}

The $\rm H\alpha$ velocity map (Fig.~6, left) shows that the gas is
rotating 
as the stars (Fig.~7, left), and that
also the stripped gas
maintains a coherent rotation for almost 60kpc downstream.
This is similar to what is observed in ESO137-001, a jellyfish galaxy in
 the Norma cluster, in which the stripped gas retains the imprint of
 the disk rotational velocity 20kpc downstream along a 30kpc tail of
 ionized gas (Fumagalli et al. 2014). This signature indicates the
 galaxy is moving fast in the plane of the sky.


The velocity
structure of the tentacles reveals that the longest tail (darkest tail
between 10 and 60 arcsec in the left top panel of Fig.~6) is separated in velocity from the
other tails and is probably trailing behind at higher negative speed
than the rest of the gas (thus the galaxy velocity vector points
  away from the observer), likely having been
stripped earlier than the other gas. This corresponds to
a region of higher velocity dispersion ($\sigma$) of the gas
(Fig.~6, right), which might indicate either that turbulent motion is setting
in, or that there is gas at slightly different velocities along the line of
sight. Inspecting the spectra, it is hard to distinguish between these two possibilities
given the faintness of the emission in this region.
Another region of high $\sigma$ is found at the southern edge of the gas 
south of the disk: a visual inspection of the spectra show that
here we are probably seeing
the superposition along the line of sight of gas at different 
location and velocities: a foreground higher velocity, and a background
lower velocity component likely stripped sooner. 

Most of the other gaseous regions and knots outside of the
disk have very low $\sigma$ (0-20/50 $\rm km \,
s^{-1}$), indicative of a dynamically cold medium.
The high velocity dispersion at the center, instead, is due to the presence of an AGN,
that will be discussed more in detail below.

In contrast with the complicated velocity structure of the gas, the
stellar component has very regular kinematics showing that the disk is
rotating unperturbed (Fig.~7, left), with a rather low velocity dispersion
(mostly between 40 and 80 $\rm km \, s^{-1}$, Fig.~7, right), as it is typical of galaxy disks. The ordered stellar
rotation, together with the regular isophotes, demonstrates that
the process responsible for the gas stripping is affecting only the
galaxy gas, and not the stars, as expected for ram pressure
stripping due to the ICM.

The difference between the stellar and gaseous velocities would
suggest a different direction 
of motion of the galaxy than the one suggested by the extended tail
(cf. Fig.~8 and Fig.~6). However, simulations indicate that the
stripped gas in the vicinity of the galaxy might not always be a
reliable indicator of the direction of motion (Roediger \& Br\"{u}ggen 2006).
The long tails indicate that the galaxy also has a significant velocity component in the 
tangential direction, on the plane of the sky.

\begin{figure}
\centerline{\includegraphics[scale=0.45]{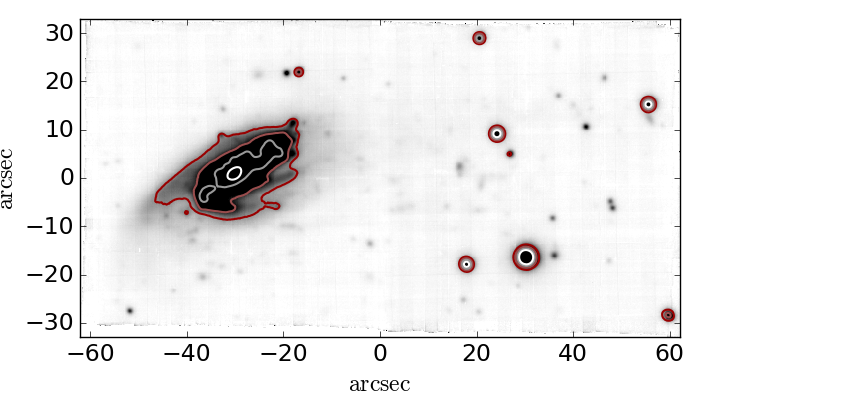}\hfill}
\caption{MUSE white image of JO206. Contours are logarithmically spaced isophotes of the continuum 
  underlying $\rm H\alpha$, thus are stellar isophotes, down to a surface brightness $2.5 \times 
  10^{-18} \rm \, erg s^{-1} cm^{-2} {\AA}^{-1} arcsec^{-2}$. Round isolated contours 
  are Galactic stars that are masked in the subsequent
  analysis. In this and all plots (0,0) is the center of the MUSE
  combined image. North is up and east is left.}
\end{figure}

\begin{figure*}
\centerline{\includegraphics[scale=0.4]{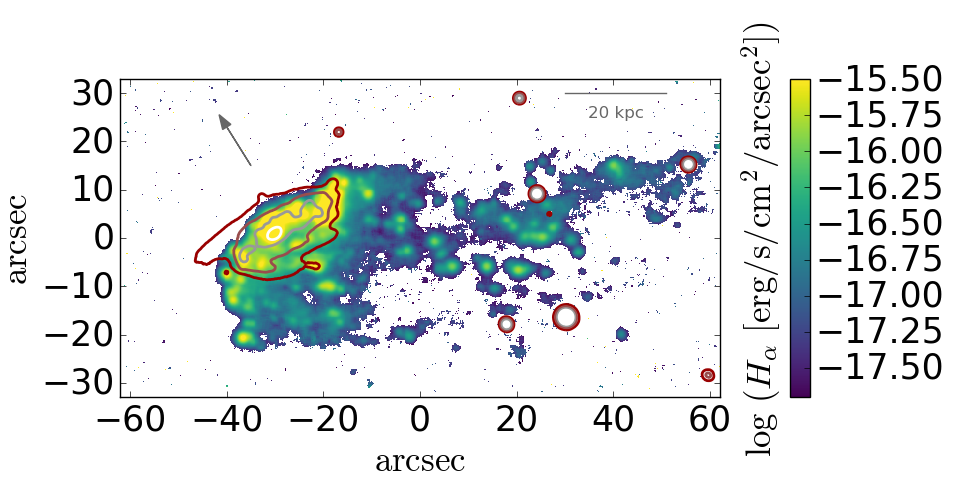}\hfill\hspace{-0.2cm}\includegraphics[scale=0.4]{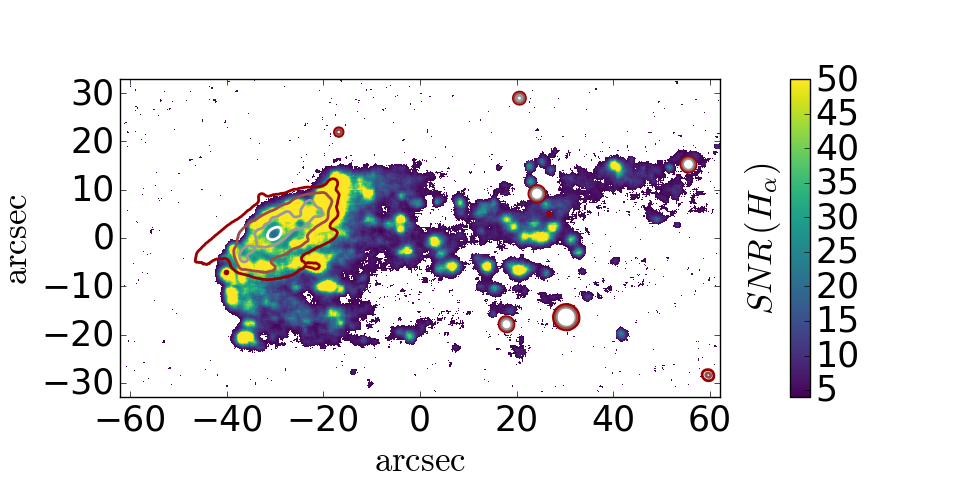}}
\caption{Left. MUSE $\rm H\alpha$ map 
  (median filtered $5\times5$ pixels) for $\rm 
  H\alpha$  S/N$>4$, uncorrected for stellar absorption and intrinsic 
  dust extinction, but corrected for Galactic extinction. The arrow indicates the direction to the BCG/X-ray center.
At the cluster redshift, 1 arcsec = 0.952 
  kpc, see scale. Contours are continuum isophotes as in Fig.~4. Round
  isolated contours are Galactic stars. Right. $\rm H\alpha$ SNR map for median filtered
  $5\times5$ pixels with S/N$>4$.}
\end{figure*}

\begin{figure*}
\centerline{\includegraphics[scale=0.4]{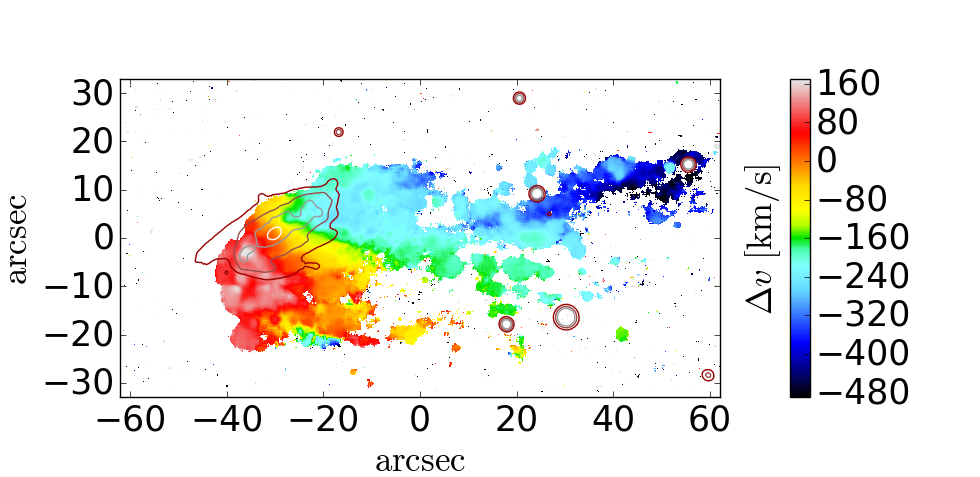}\hfill\hspace{-0.6cm}\includegraphics[scale=0.4]{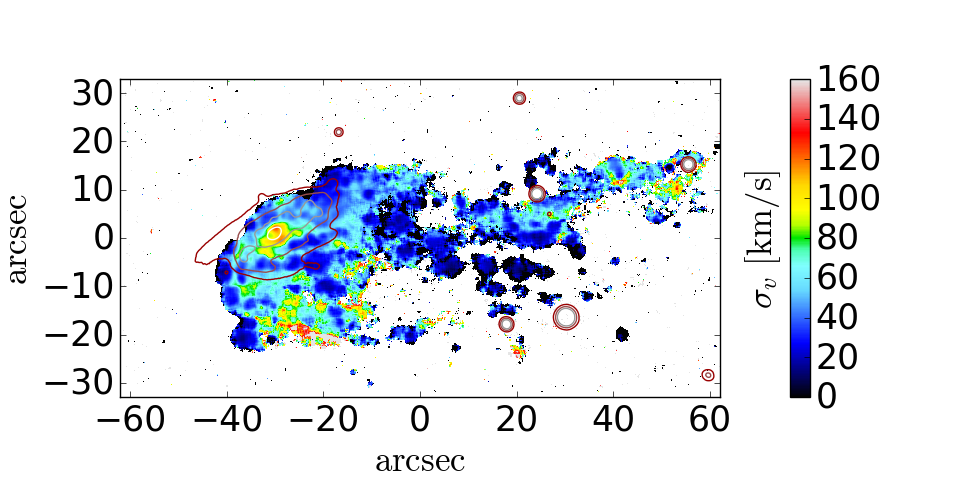}}
\centerline{\includegraphics[scale=0.4]{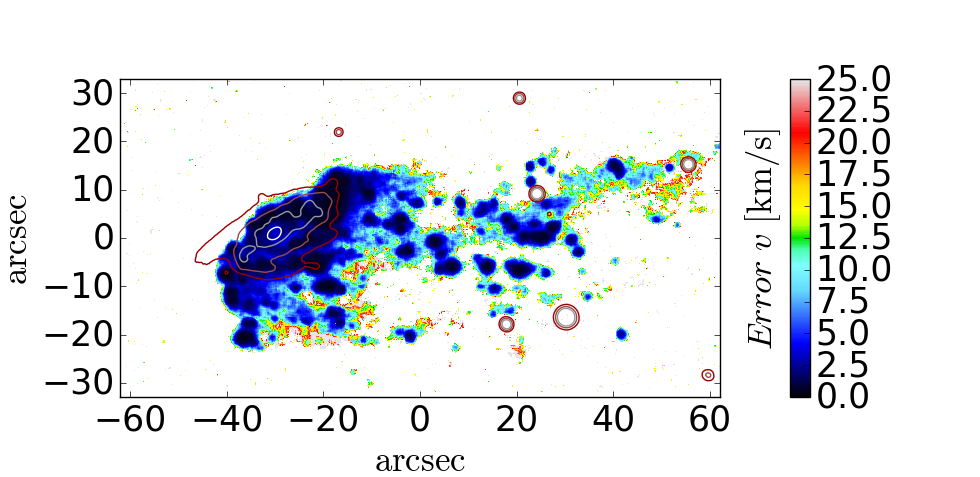}\hfill\hspace{-0.6cm}\includegraphics[scale=0.4]{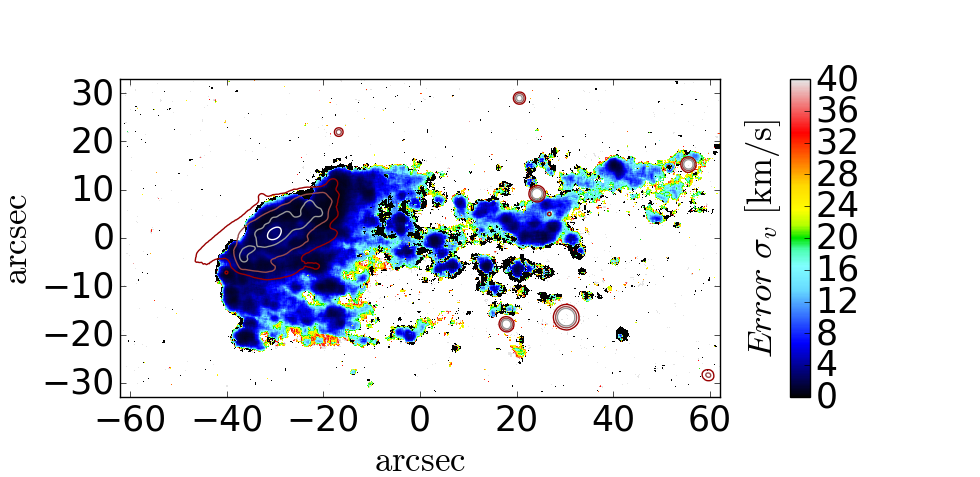}}
\caption{Top. $\rm H\alpha$ velocity (left) and velocity dispersion (right)
map for $5\times5$ spaxels with $S/N_{H\alpha}>4$. Contours are stellar
isophotes, as in Fig.~4. $v=0$ corresponds to the redshift of the
galaxy center (z=0.05133). Bottom. Corresponding error maps.}
\end{figure*}

\begin{figure*}
\centerline{\includegraphics[scale=0.4]{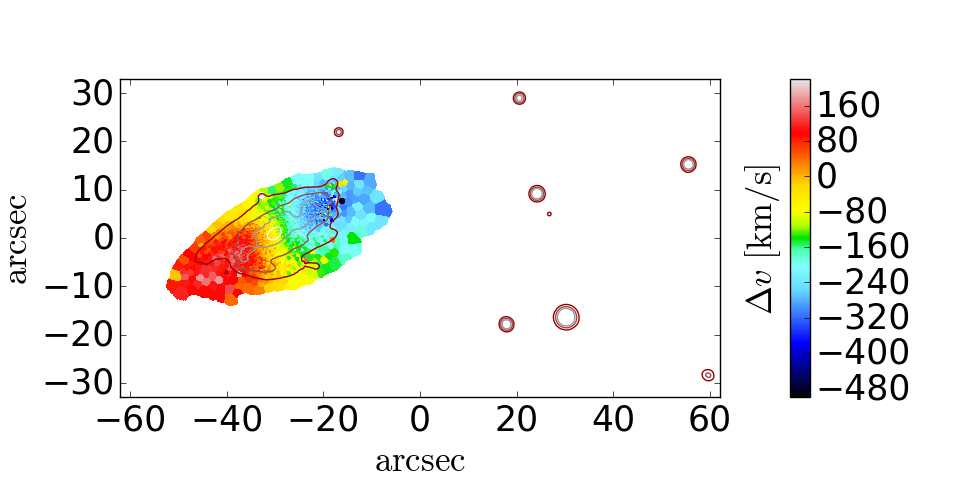}\hfill\hspace{-0.6cm}\includegraphics[scale=0.4]{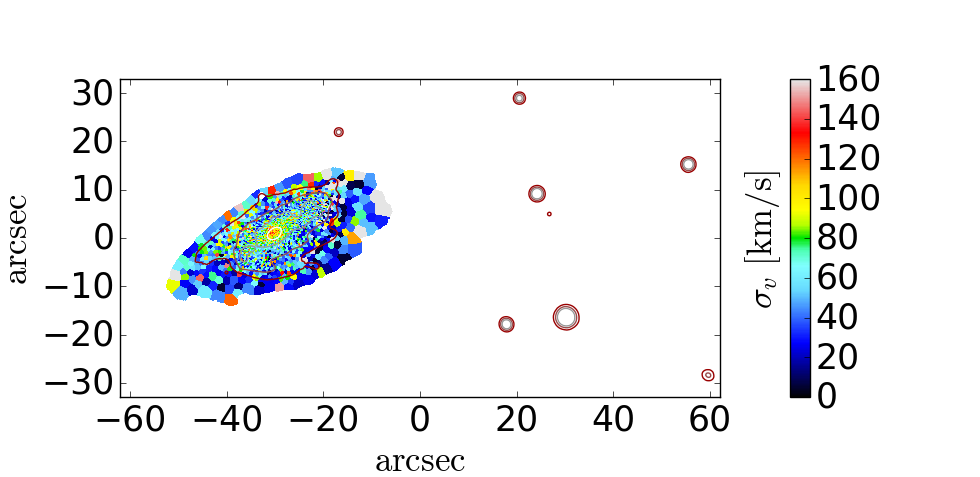}}
\caption{Stellar velocity (left) and velocity dispersion (right) map
  for Voronoi bins with S/N$>10$. Contours are stellar isophotes, as
  in Fig.~4.}
\end{figure*}

\begin{figure}
\centerline{\includegraphics[scale=0.4]{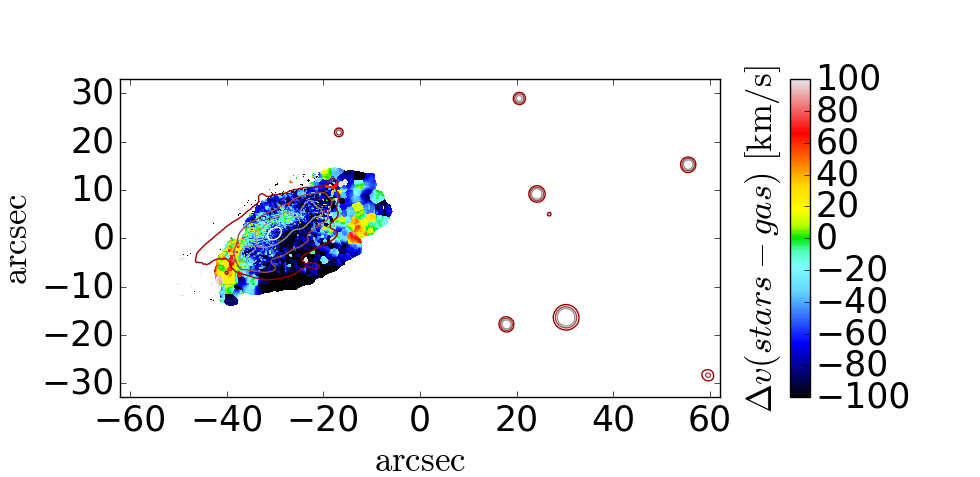}}
\caption{Velocity difference between stars and gas. Contours are
  stellar isophotes as in Fig.~4.}
\end{figure}

\subsection{Gas ionization mechanism}

The line ratio diagrams reveal that the emission in the central region is dominated by an AGN 
(Fig.~9). All three diagnostic diagrams inspected are concordant on 
this. 

The presence of an AGN in this galaxy was previously unknown. A 
posteriori, we find that JO206 coincides within the position error 
with a ROSAT bright source (EXSS 2111.2+0217, Voges et al. 1999),
hence we conclude it is an 
X-ray emitting AGN. The presence of an X-ray point source at the 
location of JO206 is also hinted by the map presented in Shang \& Scharf 
(2009). Moreover, there is a 1.4 GHz NVSS detection within 8 arcsec
from the position of JO206 (Condon et al.  1998). Using the
calibration from Hopkins et al. (2003), the 1.4Ghz flux would yield a SFR of about 12
$M_{\odot} \, yr^{-1}$, which is a factor of 2 higher than the SFR
measured from $\rm H\alpha$ (see \S 7.5).
The properties of the AGN in JO206 and in other GASP jellyfish 
galaxies will be discussed in a separate paper (Poggianti et
al. 2017b, submitted). 

Apart from the center, the [OIII]5007/$\rm H\beta$ vs
[NII]6583/$\rm H\alpha$
and vs [SII]6717,6731/$\rm H\alpha$ diagrams
show that in the rest of the disk and in all the 
extraplanar gas (including the tentacles), the emission-line ratios are 
consistent with gas being photoionized by young stars (``Star-forming''
according to Kauffmann et al. 2003 and Kewley et al. 2006) or a
combination of Star-forming and HII-AGN Composite, the latter around the
central region and in a stripe of intense H$\alpha$ brightness running
almost north-south to the north-west of the galaxy. 
The [OIII]5007/$\rm H\beta$ vs [OI]6300/$\rm H\alpha$
  diagram classifies the ionization source in these regions as LINERS,
due to the significant [OI] emission, which supports the hypothesis that
 some contribution from shocks might be present here.

What stars are responsible for the majority of the ionizing radiation?
We know that in order to produce a significant
number of ionizing photons, they must be massive stars formed
during the past $\leq 10^7$yr.
They can either be {\it new stars formed in situ, within the stripped
  gas}, or stars formed
in the galaxy disk whose ionizing radiation is able to escape to 
large distances\footnote{A third hypothesis, that the gas is ionized
  while still in the disk and then it is stripped, is unrealistic.
For a density $n=10 \, \rm cm^{-3}$, the recombination time is about
$10^4$ yr (once recombined, the decay time is negligible)
(Osterbrock \& Ferland 2006). 
Even assuming a timescale 1000 times longer ($10^7$ yr,
$n=0.01 \, \rm cm^{-3}$),
this would imply that the ionized gas had to travel 90kpc in this
time, thus at a speed of almost 9000 $\rm km \, s^{-1}$. The latter is $\sim
15$ times the cluster velocity dispersion.}.
The first hypothesis is  much more likely than the
second one for several reasons: a) the tails and knots are faint but
visible in the observed B-band light, which at the galaxy redshift should
originate from stars with no significant contribution from line
emission. In fact, the MUSE spectra at the location of $\rm H\alpha$
emission in the tentacles usually have a faint but detectable continuum;
b) as we will show in \S 7.5, the MUSE spectra in the tails can be
fitted with our spectrophotometric code with an amount of young stars
that can account for the required ionizing photons and, at the same time,
is consistent with the observed continuum level\footnote{One caveat
  is worth noting here: both arguments a) and b) might be affected by
  {\it continuum gas emission}, which is not included in SINOPSIS. The
contribution of gas emission in the continuum is currently unconstrained.};
c) if ionizing photons could travel for 90kpc without encountering any
medium to ionize, the ionized gas we see would be the only gas there
is in this area, and this is very unlikely.

Therefore, we conclude that, except for the central region
powered by the AGN, the ionization source in JO206 is mostly photoionization
by young stars.  The most likely explanation is that new massive stars
are born in situ in the stripped gas
tentacles, and ionize the gas we observe.
Our findings resemble the conclusions from
Smith et  al. (2010) regarding star formation taking place within the 
stripped gas in a sample of 13 jellyfish galaxies in the Coma
cluster and support their hypothesis that this is a widespread
phenomenon in clusters, though JO206 demonstrates that this {\sl does not
occur only in rich, massive clusters} such as Coma (see \S 7.6).



\begin{figure*}
\centerline{\includegraphics[scale=0.38]{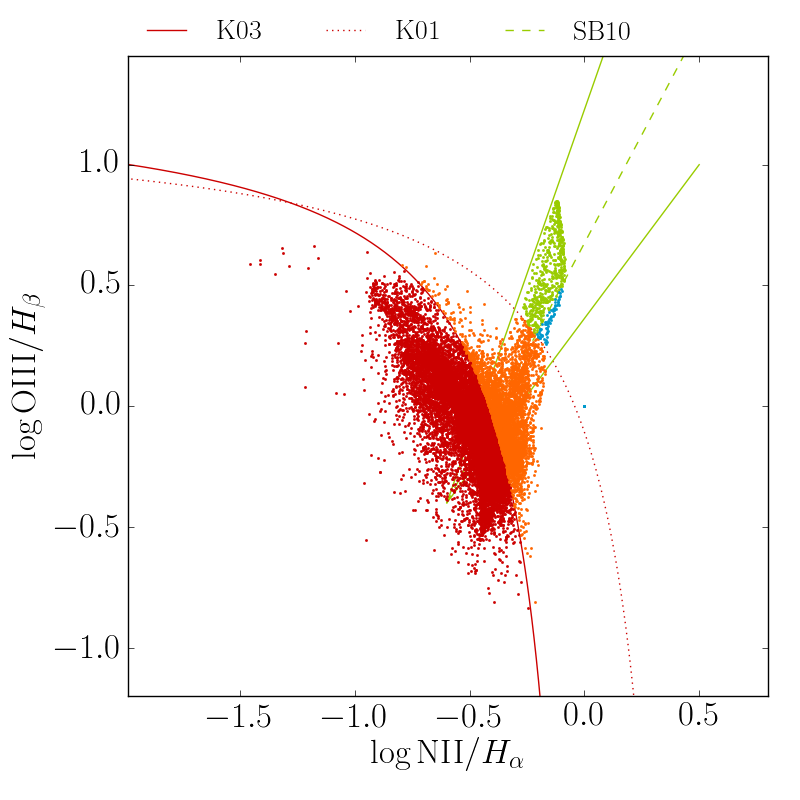}\hfill\hspace{-10cm}\includegraphics[scale=0.52]{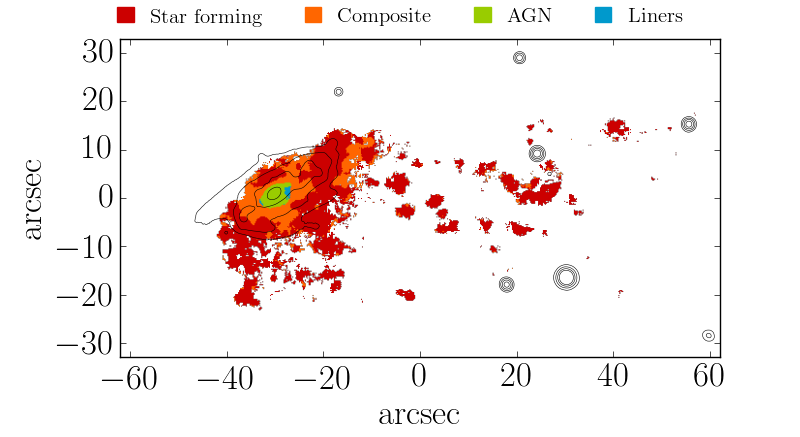}}
\centerline{\includegraphics[scale=0.38]{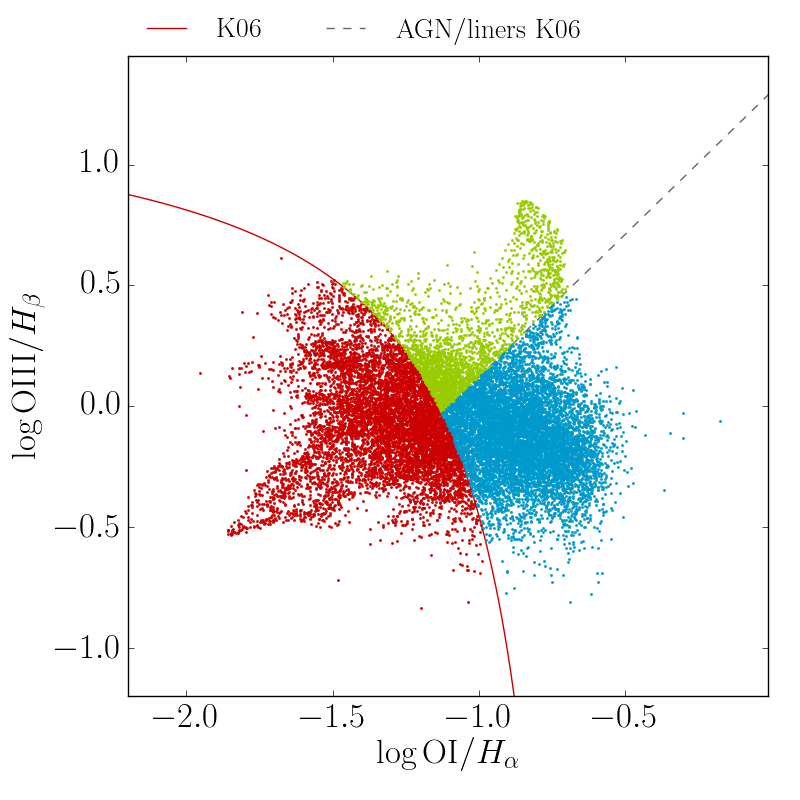}\hfill\hspace{-10cm}\includegraphics[scale=0.52]{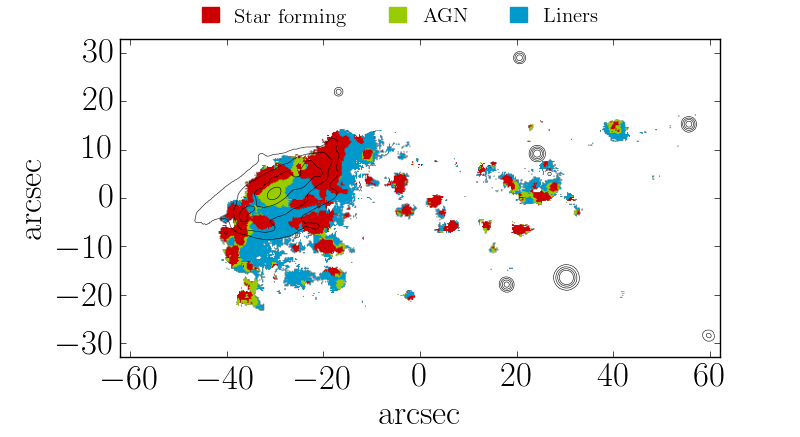}}
\centerline{\includegraphics[scale=0.38]{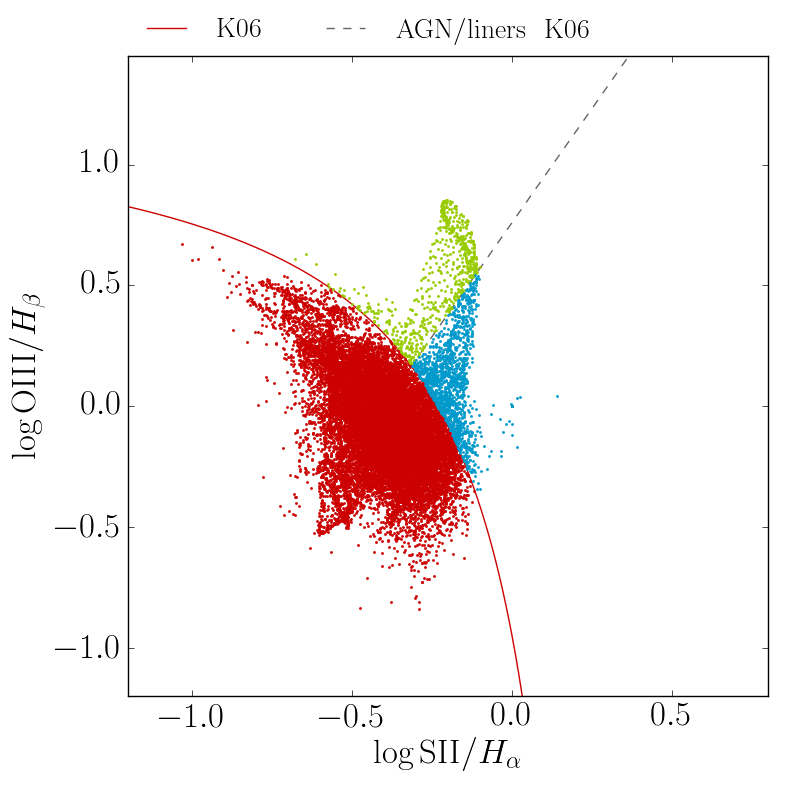}\hfill\hspace{-10cm}\includegraphics[scale=0.52]{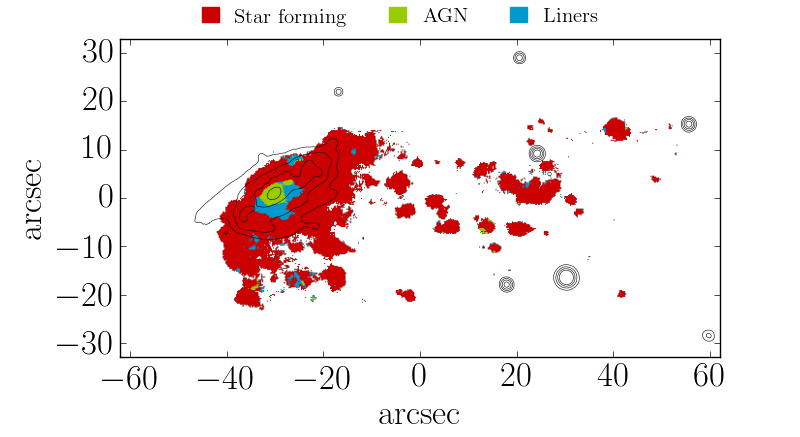}}
\caption{BPT line-ratio diagrams (left) and maps (right) for
  [OIII]5007/$\rm H\beta$ vs [NII]6583/$\rm H\alpha$ (top),
  vs [OI]6300/$\rm H\alpha$ (middle), and 
  vs [SII]6717/$\rm H\alpha$ (bottom). Lines in
the left panels are from Kauffmann et al. (2003, K03), Kewley et
al. (2001, K01, and 2006, K06) and Sharp \& Bland-Hawthorn (2010,
SB10) to separate Star-forming, Composite, AGN and LINERS. Contours
are stellar isophotes, as in Fig.~4.
}
\end{figure*}




\subsection{Dust extinction and metallicity}

The extinction map of Fig.~10 shows that the dust is not uniformly distributed,
but concentrated in knots of rather high
extinction (up to $A_V \sim 1.7$ mag), and inter-knots regions of lower
extinction values (typically 0.5-0.6 mag), with edges of
virtually no extinction. Knots of high extinction are found in the
disk but also in the tentacles, far away from the galaxy disk.
Interestingly, most of the high-extinction
regions coincide with the knots of most intense H$\alpha$ emission that
will be discussed in \S 7.4, identified in Fig.~10 by small circles.
Thus, dust appears to be concentrated in the regions with higher
H$\alpha$ brightness, hence higher SFR density, as it happens in HII regions in
normal galaxies.

\begin{figure*}
\centerline{\includegraphics[scale=0.9]{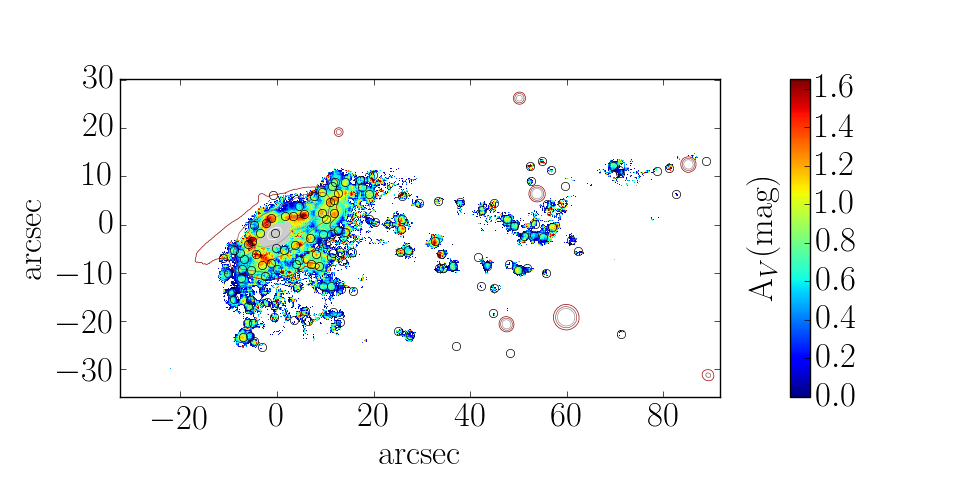}}
\caption{$A_V$ map. The small circles identify the location of the knots that are
discussed in \S 7.4. The circle radius is fixed for all knots and is not
proportional to the knot radius. The central region powered by the AGN
according to the BPT [NII]6583/$\rm H\alpha$ diagnostic has been
masked (grey area). Contours are stellar isophotes and
round isolated contours are Galactic stars, as in Fig.~4.
}
\end{figure*}

The gas metallicity varies over almost 1dex in 12+log[O/H], with the highest
metallicity regions located in the galaxy disk  (Fig.~11,
left). Interestingly, the most metal rich gas is observed $\sim 10$kpc
from the center, to the north-west of the disk. This is a star-forming
region of particularly high $\rm H\alpha$ brightness (Fig.~5, left), where the SFR per unit
area is very high, and with rather high dust extinction (Fig.~10).

The metallicity in the tentacles is intermediate to low, reaching
values as low as 12+log[O/H]=7.7-8  in the furthest regions
of the west tail and throughout the southern tail. Overall, the gas
at the end of the tentacles, likely to be the first that
was stripped
(as testified by its projected distance from
the galaxy body and/or its velocity/$\sigma$),
has a lower
metallicity. This is coherent with a scenario in which the gas in the
outer regions of the disk, which is the most metal poor, is 
stripped first, being the least bound.

Finally, the right panel of Fig.~11 shows that the ionization parameter is very
low (generally $log \, q <7$) compared to
the distribution measured in SDSS emission-line galaxies of all masses
(always $>7$, typically $7.3$, Dopita
et al. 2006). As the ionization parameter depends on several quantities
(metallicity, IMF, age of the HII region, ISM density distribution, geometry
etc), attempting an interpretation for the low values observed will require
an in-depth analysis which is beyond the scope of this paper, but will
be carried out on the whole stripped+control GASP sample.

\begin{figure*}
\centerline{\includegraphics[scale=0.42]{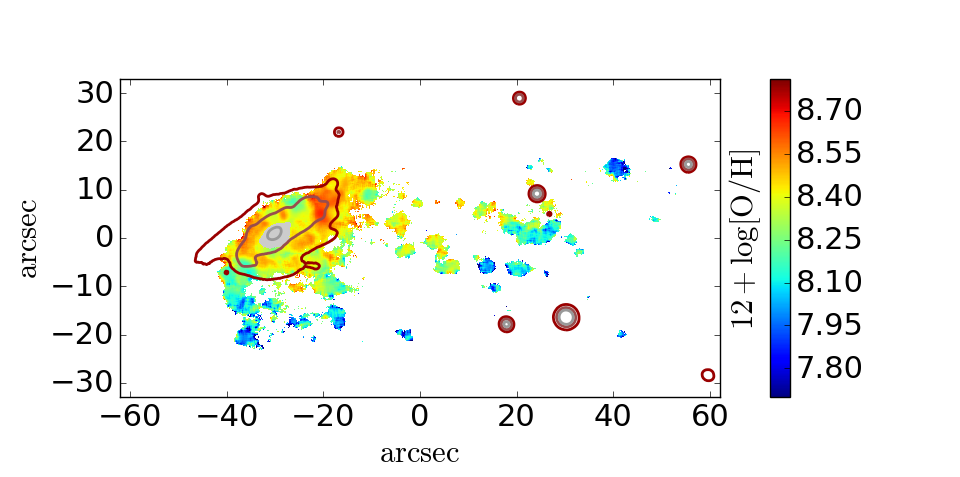}\hfill\hspace{-0.6cm}\includegraphics[scale=0.42]{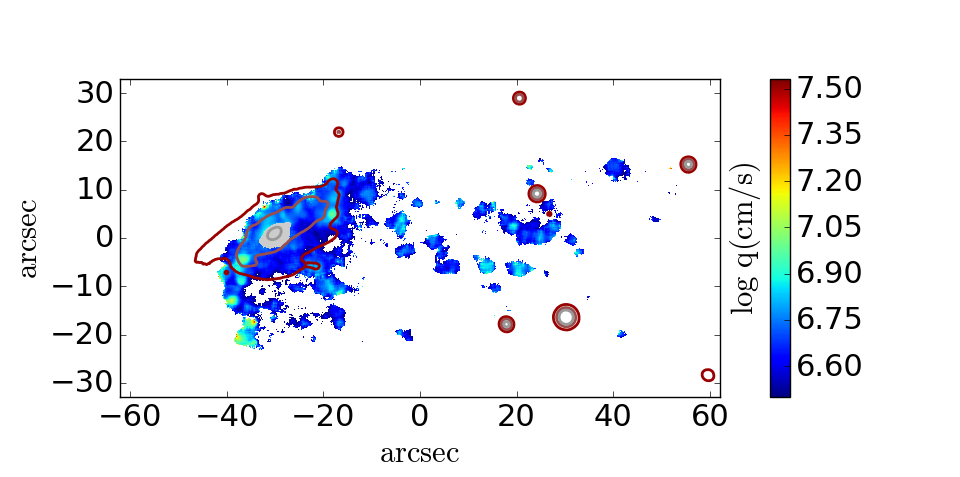}}
\caption{Metallicity (left) and ionization parameter (right) map. The central region powered by an AGN
according to the BPT [NII]6583/$\rm H\alpha$ diagnostic has been
masked (grey area). Contours are stellar isophotes as in Fig.~4.}
\end{figure*}


\subsection{Star-forming knots}
We identify 139 individual knots in the JO206 H$\alpha$ image, as
described in \S 6.4 and shown in Fig.~12. These are regions of high
H$\alpha$ surface brightness, typically log ($\rm H\alpha [erg s^{-1}
cm^{-2} arcsec^{-2}]) \sim -15.5$. 
Figure~13 shows that, except for the central H$\alpha$ knot dominated
by the AGN and a few  
small surrounding knots powered by a Composite source, 
the spectra of all the other knots are consistent with photoionization from 
young stars. 

We note that MUSE revealed $\rm H\alpha$ knots 
also in the tails of ESO137-001 (Fossati et 
  al. 2016) and these authors conclude these are HII regions formed in 
  situ, as we find for JO206. In situ condensation of stripped gas is 
  also found by Yagi et al. (2013) in the star-forming regions around 
  NGC4388 in Virgo, and other extragalactic HII regions are known in 
  Virgo (Cortese et al. 2004, Gerhard et al. 2002). 
In contrast,
Boselli et al. (2016) conclude that NGC4569,
a spectacular jellyfish in Virgo studied with narrow-band $\rm  H\alpha$+[NII] imaging,
lacks star-forming regions in the tail and therefore suggest that 
the gas is excited by mechanisms other than photoionization 
(e.g. shocks, heat conductions etc.). 
It will be interesting to 
understand how common are HII regions in jellyfish tails once 
the whole GASP sample is available, see for example the galaxies 
JO201 in Paper II and JO204 in Paper V.


\begin{figure}
\centerline{\includegraphics[scale=0.42]{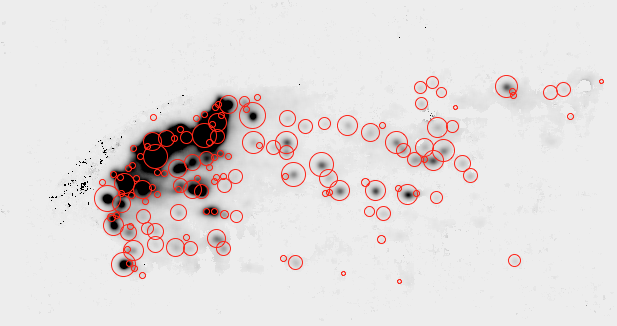}}
\caption{$\rm H\alpha$ flux map (black) with $\rm H\alpha$ knots as red circles.}
\end{figure}


\begin{figure*}
\centerline{\includegraphics[scale=0.31]{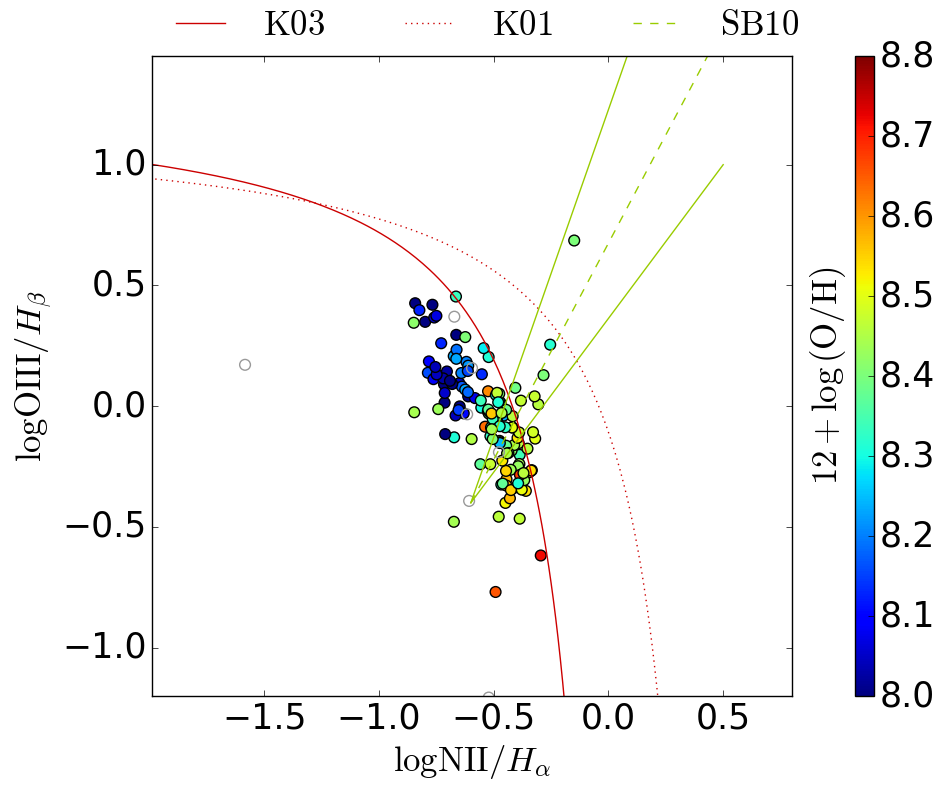}\hfill\hspace{-0.5cm}\includegraphics[scale=0.5]{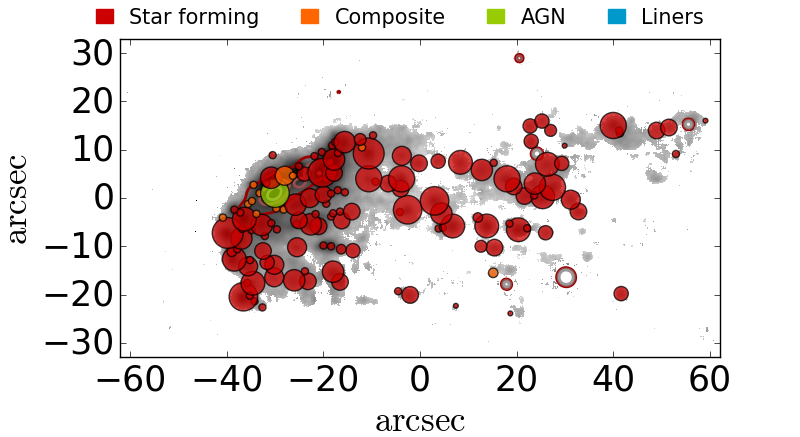}}
\caption{Left. BPT diagram of the $\rm H\alpha$ knots, color-coded by
  metallicity. Right. Spatial distribution of the knots, color-coded by
  ionization source. Here the radius of the circle corresponds to the
  radius of the knot (see sec.~6.4). The grey shaded area represents
  the $\rm H\alpha$ image.}
\end{figure*}

 
%

In our MUSE data the metallicity of the gas in the knots varies with knot location (Fig.~14), and
traces the spatially resolved metallicity shown in Fig.~11.
The knots south of the disk, on the southern side of the main tail and those at
the highest distances in the tail to the west are mostly metal poor
(12+log(O/H)=8.0-8.2).
The majority of the rest of the knots have intermediate metallicities
(8.3-8.4), while the most metal rich significant knots are located along the disk
north-west of the galaxy center.

Thus, the numerous knots we observe in JO206 appear to be giant HII regions and
complexes.
We estimate the
ongoing SFR in each knot as described in \S 6.3. The sum of the SFR
in all knots is 5.2 $M_{\odot} \, yr^{-1}$. The $\rm H\alpha$ luminosity of
the only knot powered by the AGN corresponds to $\sim 1 M_{\odot} \, yr^{-1}$, thus the total
SFR in all blobs excluding the AGN is $\sim 4.2 M_{\odot} \, yr^{-1}$.
The SFR distribution of the knots is shown in Fig.~15. It ranges from
a minimum of $\sim 10^{-4}$ to a maximum of 0.8 $M_{\odot} \, yr^{-1}$
per individual knot, and the average is about 0.01 $M_{\odot} \, yr^{-1}$.
The SFR determination has several important caveats.
These values are derived assuming a Chabrier IMF,
but the true shape of the IMF in the knots is 
unconstrained. Even assuming a known IMF, at such low values of
SFR of individual knots, the stochasticity of the 
IMF sampling can be important and will be the subject of a subsequent study.

The distributions of the ionized gas densities of the individual knots are shown in Fig.~15. 
Of the 138 knots with no AGN, 91 have a [SII]6716/[SII]6732 ratio in the
range where the density calibration applies (see \S6.3). The remaining knots have 
ratio values larger than 1.44, which 
suggests that their density is below 10 $\rm cm^{-3}$.
Figure~15 shows that the most of the measured densities  are between
10 and 100 $\rm cm^{-3}$, with a median of 28 $\rm
cm^{-3}$\footnote{We have checked the spatial distributions of the
  knots with a density estimate (not shown) and they are distributed both in the
  disk and in the tails, tracing all the regions with $\rm H\alpha$
  emission. A detailed study of the physical properties of the
  individual HII regions in GASP galaxies will be the subject of a
  forthcoming paper.}.
For these, we derive the ionized gas mass
following eqn.~3 to derive the knot gas mass distribution also shown in Fig.~15.
Most of these knots have masses in the range $10^4-10^{6.5}
M_{\odot}$, with a median of $\sim 1.5 \times 10^5 M_{\odot}$.
Summing up the gas mass in these knots we obtain 
$1.7 \times 10^8 \, M_{\sun}$, which represents a hard lower limit to the 
total ionized gas mass, given that the contributions of the knots with
no density estimate and of the diffuse line emission are not taken
into account. However, considering that the knots contributing to this
mass estimate already account for about half of the total $L_{\rm H\alpha}$, the
total mass value given above is probably of the order of the true
value.

Once star formation will be exhausted in JO206 blobs, they will
  probably resemble the UV ``fireballs'' in the tail of IC3418, a dIrr
galaxy in Virgo with a tail of young stellar blobs ($<$400 Myr, Fumagalli et
al. 2011, Hester et al. 2010, Kenney et al. 2014).



%

%

\begin{figure*}
\centerline{\includegraphics[scale=0.42]{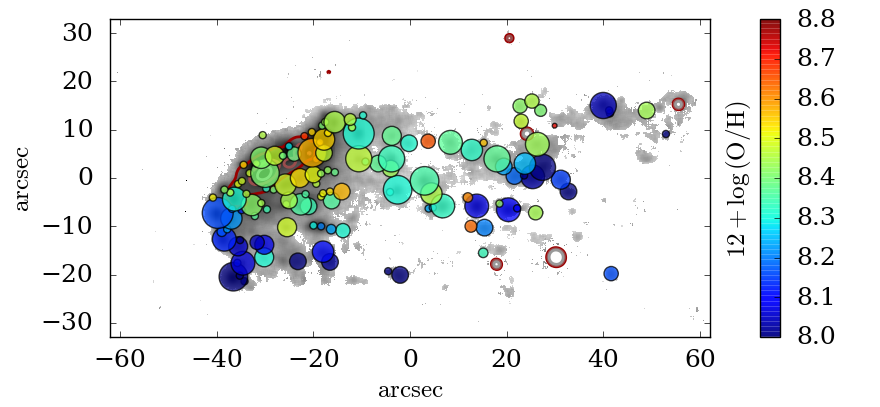}\hfill\hspace{-0.3cm}\includegraphics[scale=0.42]{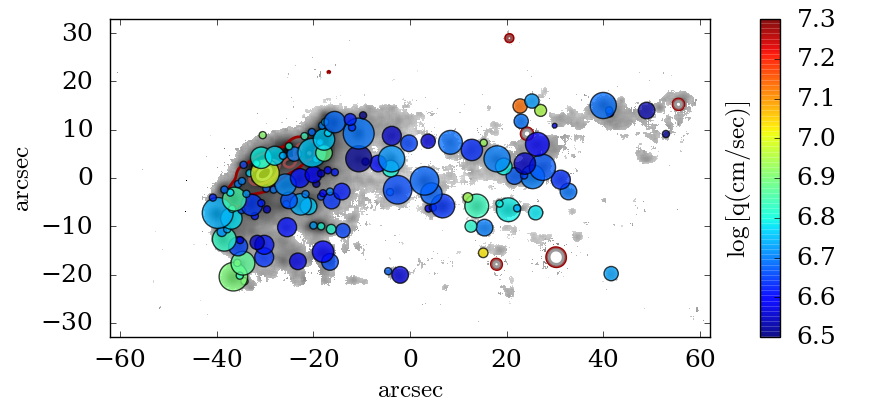}}
\caption{Metallicity (left) and $q$ ionization parameter (right) of 
  the knots. The grey shaded area represents the $\rm H\alpha$ image.}
\end{figure*}

\begin{figure*}
\centerline{\includegraphics[scale=0.42]{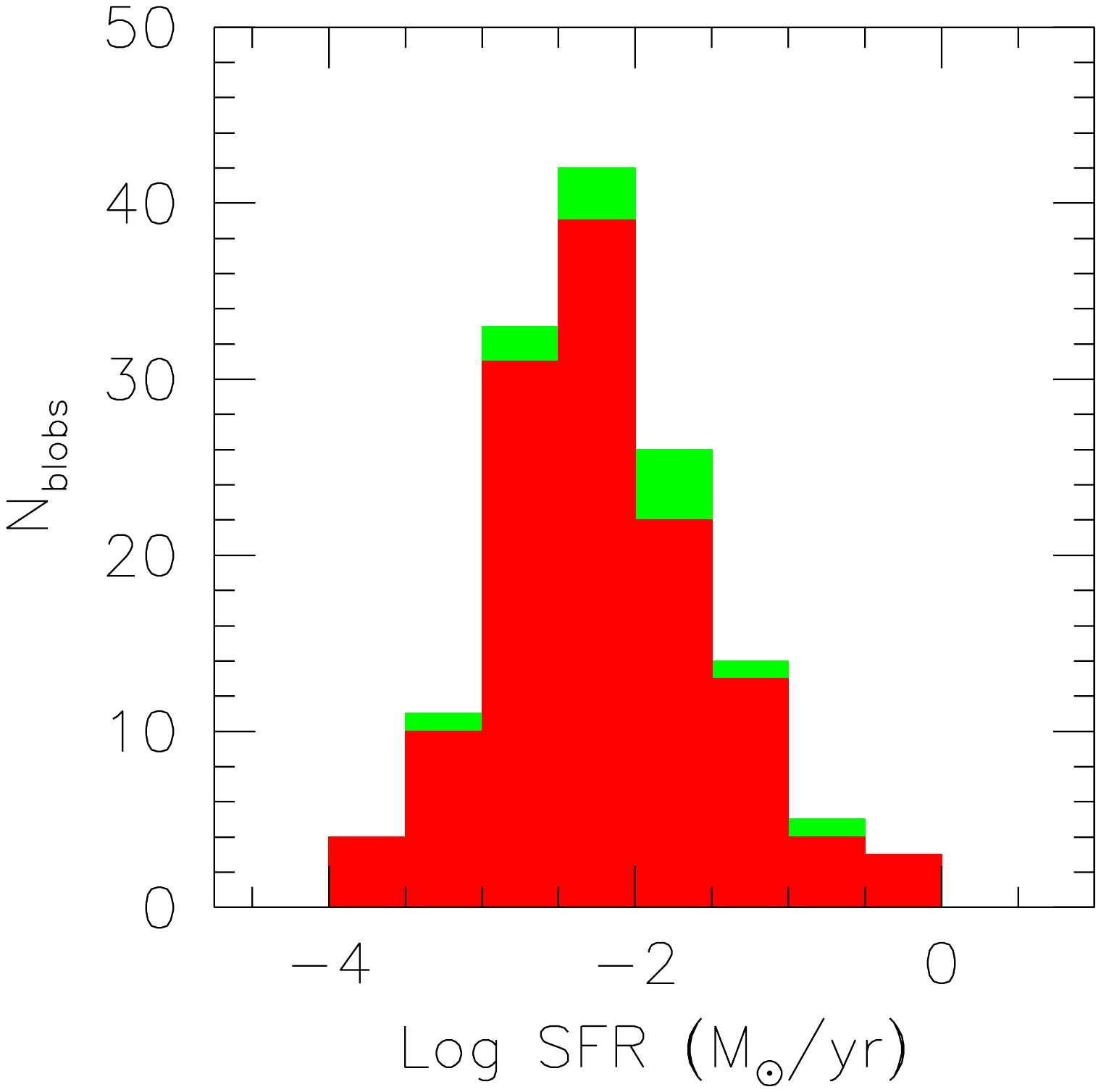}\hfill\hspace{-2.5cm}\includegraphics[scale=0.42]{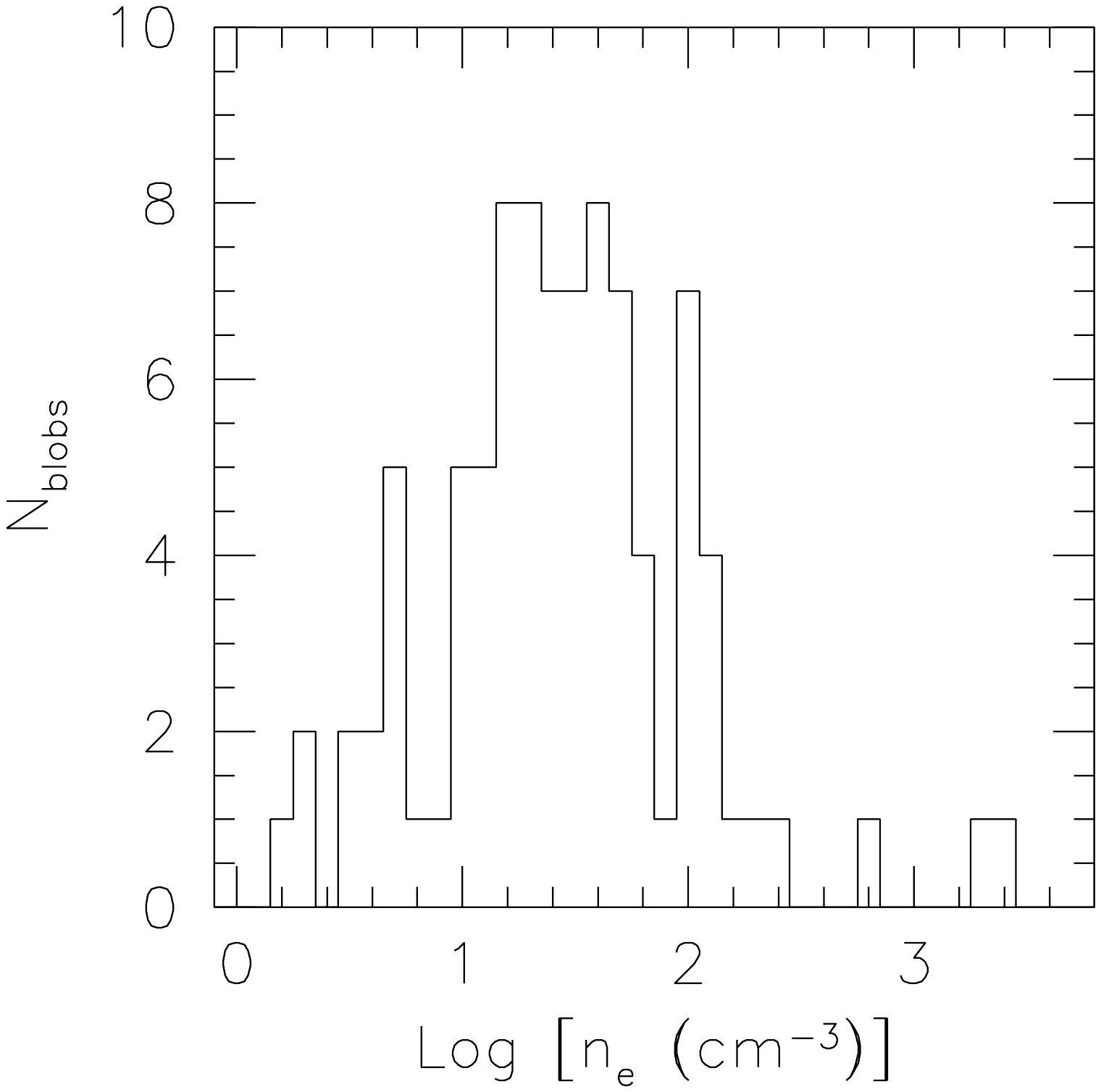}\hfill\hspace{-2.5cm}\includegraphics[scale=0.42]{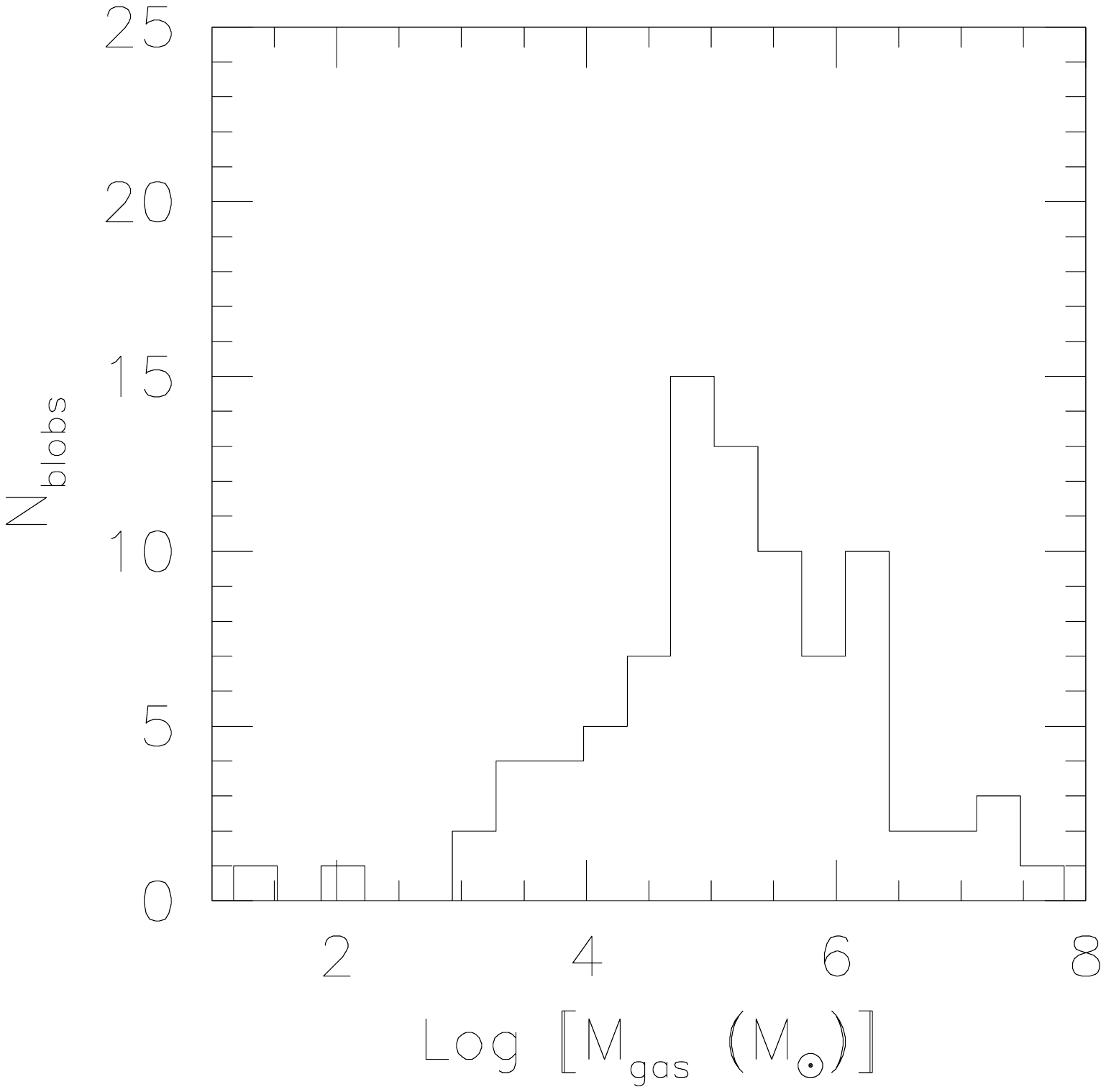}}
\caption{Left. SFR distribution of the ``Star-forming'' (red) and
  ``Composite'' (green) knots. The AGN-powered central knot has been
  excluded. Center. Distribution of gas densities of individual
  knots. Other $\sim 45$ knots not included in the plot have densities
  probably lower than the range shown here (see text).
Right. Distribution of ionized gas masses of the individual
knots plotted in middle panel. }
\end{figure*}



\subsection{Stellar history and star formation}
The total SFR computed from the dust- and absorption-corrected $\rm H\alpha$ luminosity is 6.7
$M_{\odot} \, yr^{-1}$, of which 4.3/5.9 $M_{\odot} \, yr^{-1}$ within the inner/outer continuum
isophotes. The SFR outside of the galaxy main body is therefore 
$\sim 1-2.5 M_{\odot} \, yr^{-1}$.
Subtracting the contribution from the regions that are
classified as AGN or LINERS from the BPT diagram, the total SFR
remains 5.6 $M_{\odot} \, yr^{-1}$.
 
The spatially resolved stellar history is reconstructed from SINOPSIS,
which allows us to investigate how many stars were formed at each
location during four logarithmically-spaced periods of time
(Fig.~16).

The ongoing star formation activity  (stars formed during the last $2 \times 10^7
yr$, top left panel in Fig.~16), is very intense along the east
side of the disk, but is essentially absent in the easternmost
stellar arm where the gas has already been totally stripped (cf. Fig.~5)\footnote{It should be kept in mind that in the
  central region, where the
  gas is ionized by the AGN (cf. Fig.~8), the ongoing SFR is
  overestimated by SINOPSIS.}.
Ongoing star formation is also present  throughout
the stripped gas, including the tails far out of the galaxy, with
knots of higher than average SFR spread at different locations. 

The recent star formation activity (between $2 \times 10^7 yr$ and
$5.7 \times 10^8 yr$, top right panel) has a slightly different spatial distribution
compared to the youngest stars: recent star formation was present also
in the easternmost galaxy arm, a wide region of SF activity was
present to the west of the galaxy body (X coordinates=-20 to 0), and
the SFR in the tentacles was more rarefied.
In agreement with this, inspecting the spectra in the easternmost galaxy arm
shows typical post-starburst (k+a, Dressler et al., 1999, Poggianti et al. 1999) features, with no emission
lines and extremely strong Balmer lines in absorption (rest frame $\rm
H\beta \sim 10 \AA$).

The distribution of older stars ($> \sim 6 \times 10^8 yr$) is
drastically different, being mostly confined to the main galaxy body
(two bottom panels of Fig.~16).

The different spatial distributions of stars of different ages
indicate that the star formation in the stripped gas was ignited
sometime during the last $\sim 5 \times 10^8$yr. 

As a consequence of the spatially-varying star formation history, the
stellar luminosity-weighted age varies with position
(Fig.~17). Overall, there is a clear age gradients from older to
younger ages going from east to west\footnote{Again, the reader should
  remind that the central region is contaminated by the AGN.}, 
with a few noticeable exceptions: on the galaxy disk,
there are some regions of very young ages, corresponding to the high
SFR density regions in Fig.~16, that coincide with very
intense H$\alpha$ surface brightness (cf. Fig.~5) and high metallicity
(Fig.~10) whose emission is powered by star formation (Fig.~9). 
The tails have low LW ages, as expected given the star formation
history of Fig.~16.

While the distribution of recent star formation is driven by the
stripping of the gas, the stellar mass density distribution is
dominated by the old stellar generations (Fig.~17, right). The mass density
in the tails and in the stripped gas in general is very low, more than
two orders of magnitude lower than in the galaxy disk (where the
density is highest), and about an order of magnitude lower than in the
outer regions of the disk.



\begin{figure*}
\centerline{\includegraphics[scale=0.43]{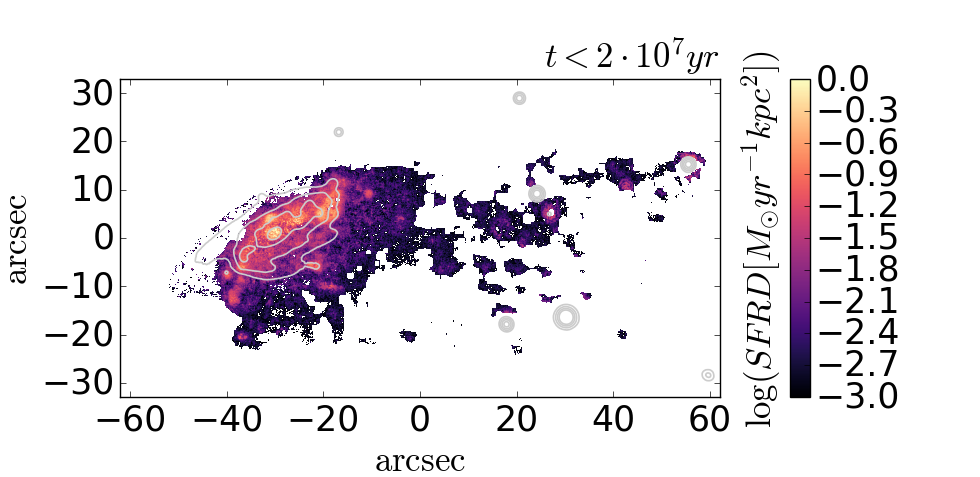}\hfill\hspace{-0.6cm}\includegraphics[scale=0.43]{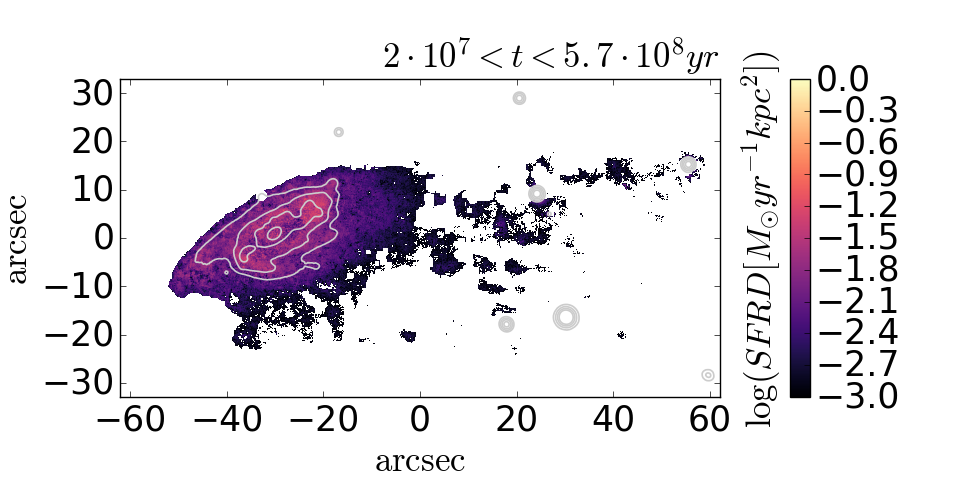}}
\centerline{\includegraphics[scale=0.43]{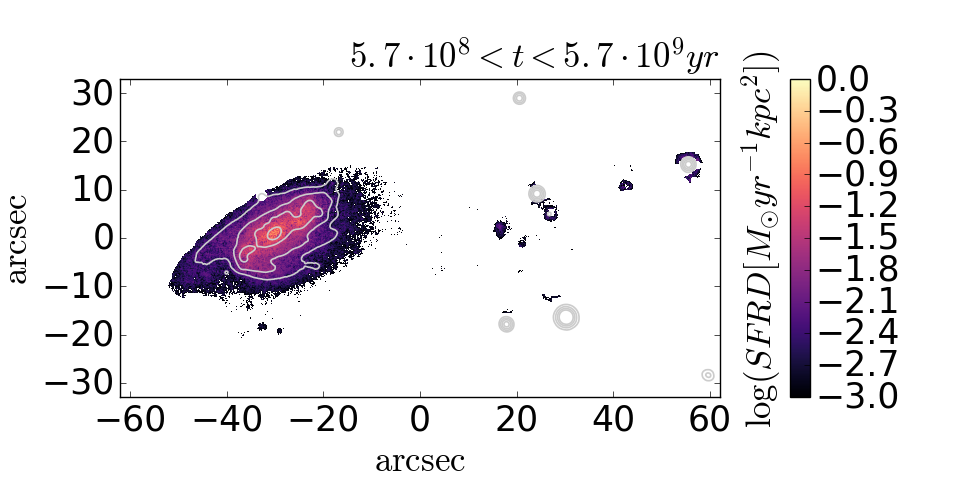}\hfill\hspace{-0.6cm}\includegraphics[scale=0.43]{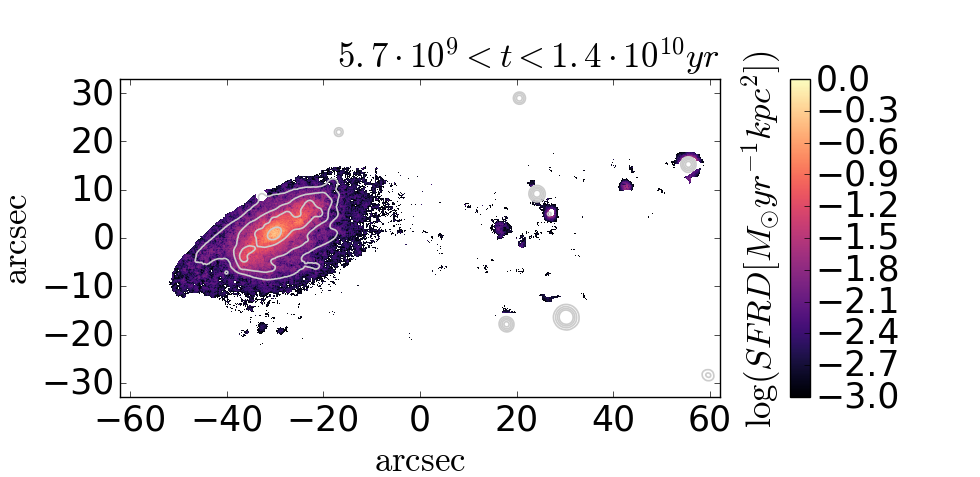}}
\caption{Stellar maps of different ages, illustrating the average star
formation rate per $\rm kpc^2$ during the last $2 \times 10^7 yr$ (top
left), between $2 \times 10^7 yr$ and $5.7 \times 10^8 yr$ (top
right), $5.7 \times 10^8 yr$ and $5.7 \times 10^9 yr$ (bottom left)
and $>5.7 \times 10^9 yr$ ago (bottom right). 
Contours in all panels are continuum isophotes as in Fig.~4.
}
\end{figure*}

\begin{figure*}
\centerline{\includegraphics[scale=0.42]{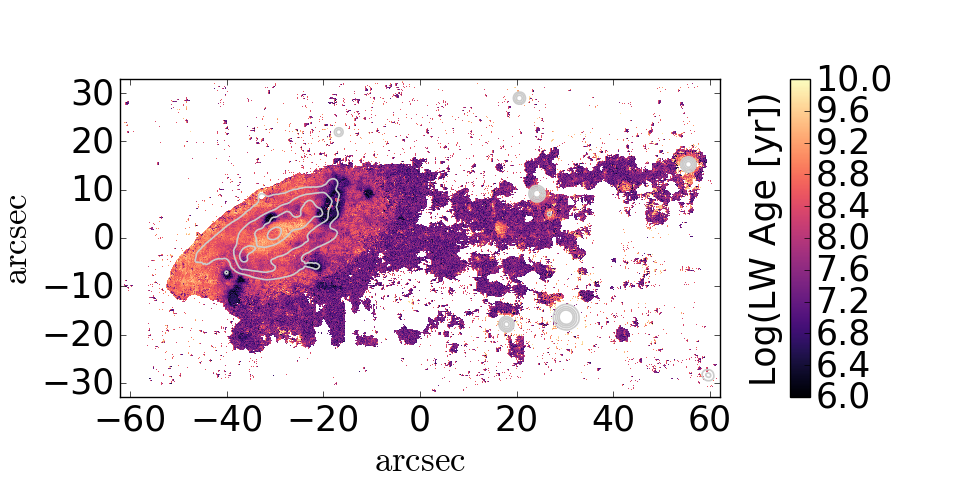}\hfill\hspace{-0.6cm}\includegraphics[scale=0.42]{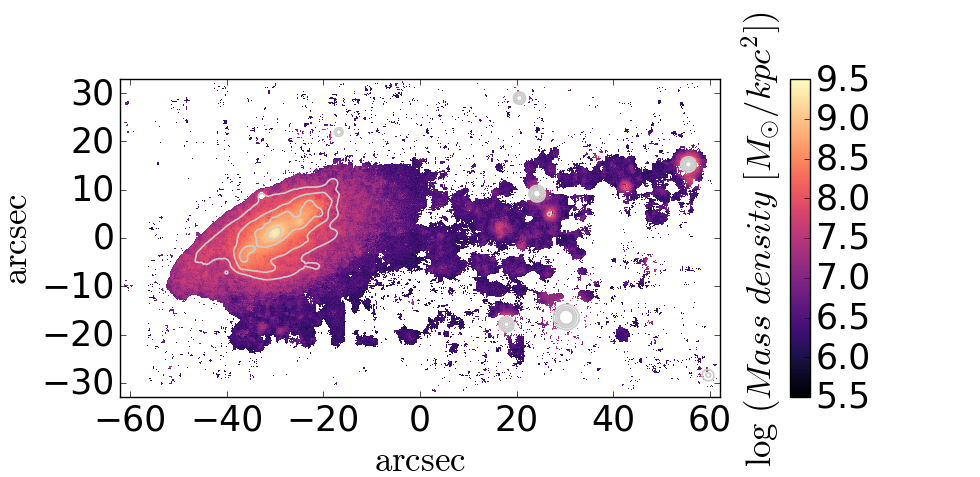}}
\caption{Left. Map of luminosity-weighted stellar age. Right: Stellar
  mass density map. Contours in both panels are continuum isophotes as in Fig.~4.}
\end{figure*}



\subsection{JO206 environment}
IIZW108 is a poor galaxy cluster with an X-ray luminosity 
$L_X= 1.09 \times 10^{44} \, \rm erg \, s^{-1}$ (ROSAT 0.1-2.4 keV, Smith et
al. 2004). In the literature (e.g. in SIMBAD and NED), it is commonly
refereed to as a {\it galaxy group}, for its modest X-ray luminosity
and temperature ($T_X=3.93\pm 0.1$ keV, Shang \& Scharf 2009) and
optical richness. According to previous studies, IIZW108 has 
little intracluster light and is 
undergoing major merging in its central regions, with four galaxies now in the process of building up 
the BCG (Edwards et al. 2016, see also Fig.~3). 
The only velocity dispersion estimates for this cluster
are from WINGS and OMEGAWINGS: 549$\pm$42 $\rm km/s$ (Cava et al. 2009), and
revised values of 611$\pm$38 $\rm km/s$ based on 171 spectroscopic members 
(Moretti et al. 2017) and 545/513$^{+37}_{-35} \rm km/s$ based on 179 spectroscopic
members including/excluding galaxies in substructures (Biviano et al. in
prep.). 
Cluster mass and radius are estimated from the dynamical analysis 
of Biviano et al. (in prep.) with the MAMPOSSt technique (Mamon et
al. 2013) to be $M_{200} = 1.91^{+0.96}_{-0.45} \times 10^{14} \,
M_{\sun}$ and $R_{200}=1.17^{+0.17}_{-0.10} \rm \,
Mpc$.\footnote{$R_{200}$ is defined as the projected radius delimiting
a sphere with interior mean density 200 times the critical density of
the Universe, and it is a good approximation of the cluster virial radius.}

The dynamical analysis of IIZW108 confirms that it
has a highly significant substructure in the central region,
and shows evidence for a few additional, less significant substructures,
distributed from the north-east to the south-west of the cluster, as 
shown in Fig.~18
(see also Biviano et al. in prep.).
However, there is no evidence for JO206 to reside in any of these
substructures. On the contrary, this galaxy appears to have fallen 
recently into
the cluster as an isolated galaxy.

JO206 is located in the most favorable conditions
for ram pressure stripping within the cluster: it is
at a small projected cluster-centric radius ($r_{cl} \sim 0.3 R_{200}$
from the BCG), and it has a high differential velocity
with respect to the cluster redshift ($\Delta v_{cl} \sim 800 km s^{-1} 
\sim 1.5 \sigma_{cl} $) (Fig.~18).
We can compute the expected ram-pressure on JO206 by the ICM as
$P_{ram} = \rho_{ICM} \times v_{cl}^{2}$ (Gunn \& Gott 1972), where
$\rho_{ICM}(r_{cl})$ is the radial density profile of the ICM.
As we do not have a good estimate of $\rho_{ICM}(r_{cl})$ for IIZW108, 
we used the well-studied Virgo cluster
as a close analogue (the clusters have very similar mass), assuming a 
smooth static ICM:

\begin{equation}
\rho_{_{ICM}}(r_{cl})=\rho_0 
\left[1+\left(\frac{r_{cl}}{r_c}\right)^2\right]^{-3\beta/2},
\end{equation}

with core radius $r_c = 13.4 kpc$ , slope parameter $\beta = 0.5$, 
and central density $\rho_{0} = 4 \times 10^{2} cm^{-3}$
(the same values used in Vollmer et al. 2001). 
At the projected $r_{cl}$ and line-of-sight velocity of JO206,  we
get a lower limit to the pressure:
\begin{equation}
P_{ram} = 6\times10^{-14} N m^{-2}
\end{equation}
We then compare the ram pressure of IIZW108 with the anchoring force of 
an idealized disk galaxy with the properties of JO206.
The anchoring force of a disk galaxy $\Pi_{gal} = 2 \pi G \Sigma_{g} 
\Sigma_{s}$ is a function of
the density profiles of the stars and the gas components ($\Sigma_{g}$ 
and $\Sigma_{s}$ respectively),
that can be expressed as exponential functions:

\begin{equation}
\Sigma=  (\frac{M_{d}}{2 \pi r_d^2})  e^{-r'/r_d},
\label{sigma0}
\end{equation}

where $M_{d}$ is the disk mass, $r_d$ the disk scale-length and r the 
radial distance from the center of the galaxy.
For the stellar component of JO206 we adopted a disk mass 
$M_{d,stars}=8.5\times10^{10}M_{\odot}$,
and a disk scale-length $r_{d,stars} = 5.73 $~kpc, obtained by fitting 
the light profile of the galaxy with GASPHOT (D'Onofrio et al. 2014).
For the gas component we assumed a total mass $M_{d,gas} = 0.1 \times 
M_{d,stars}$,
and scale-length $r_{d,gas} = 1.7 \times r_{d,stars}$ (Boselli 
\& Gavazzi 2006).

At the centre of the galaxy ($r=0$) the anchoring force is too
  high for stripping to happen ($\Pi_{gal}(r=0) 
\simeq 10^{-11} N m^{-2}$).  The condition for stripping
($P_{ram}$/$\Pi_{gal} > 1$) is only met at a radial distance of 
$r\simeq20 kpc \sim 2 \times r_{d,gas}$ from the centre of the disk,
which coincides very well with the  ``truncation radius'' ($r_{t}$)
measured from  the extent of $\rm H{\alpha}$ emission (see
e.g. Fig.~5). At $r=r_{t}$ the anchoring force drops to 
\begin{equation}
\Pi_{gal}(r=r_{tr})\simeq 4\times 10^{-14} N m^{-2} \sim 0.7 \times P_{ram}
\end{equation}
and the fraction of remaining gas mass can be computed as:   
\begin{equation}
f=1+\left[e^{-r_t/r_d}\left(\frac{-r_t}{r_d} -1\right)\right] \label{eq:frac}
\end{equation}



This simplified calculation yields a gas mass
fraction lost to the ICM wind of 15\%. 
This amount of stripping happens at the combination of clustercentric distances and 
velocities  shown by the  blue lines in the right-hand panel of Fig.~18, where the condition
$ P_{ram} \simeq \Pi_{gal}(r = r_{tr})$ is met. The infalling galaxy has been approaching 
the cluster core moving from the right hand side of the plot to the left, and gaining absolute velocity (see e.g. Fig.~6 in Yoon et al. 2007).  
At the time of observation the galaxy is crossing the 15\% stripping line, and it will continue to get stripped if it reaches lower $r_{cl}$. 


We note that, although our calculation for JO206 supports ongoing
ram-pressure stripping,
there are several caveats to be considered
(see discussion in e.g. Kenney et al. 2004 and Jaff\'e et al. 2015).
First, the three-dimensional position and velocity of the galaxy are 
unknown,
we use projected values. Second, in our calculations we assumed an 
idealized exponential disk
interacting face-on with a static and homogeneous ICM.
Simulations have shown however, that the intensity of ram-pressure 
stripping can be enhanced in cluster mergers
(Vijayaraghavan \& Ricker 2013) due to the presence of higher 
density clumps or shock waves,
and that its efficiency varies with galaxy inclination 
(Abadi et al. 1999, Quillis et al. 2000, Vollmer et al. 2001).
Finally, given the current lack of a direct measurement of the cold gas 
component of JO206,
we used the extent of $\rm H\alpha$ to estimate the amount of gas stripped.
It is yet to be tested with approved JVLA observations whether HI is 
more truncated than $\rm H\alpha$ in this jellyfish galaxy.

JO206 is therefore an example of a high mass galaxy undergoing strong
ram pressure stripping in a poor, low-mass cluster. JO206 is not
  the only known jellyfish with these characteristics: NGC 4569 is also
a quite massive ($10^{10.5}
  M_{\odot}$) galaxy in a $\sim 10^{14} M_{\odot}$ cluster (Virgo)
(Boselli et al. 2016).



\begin{figure*}
\centerline{\includegraphics[scale=0.55]{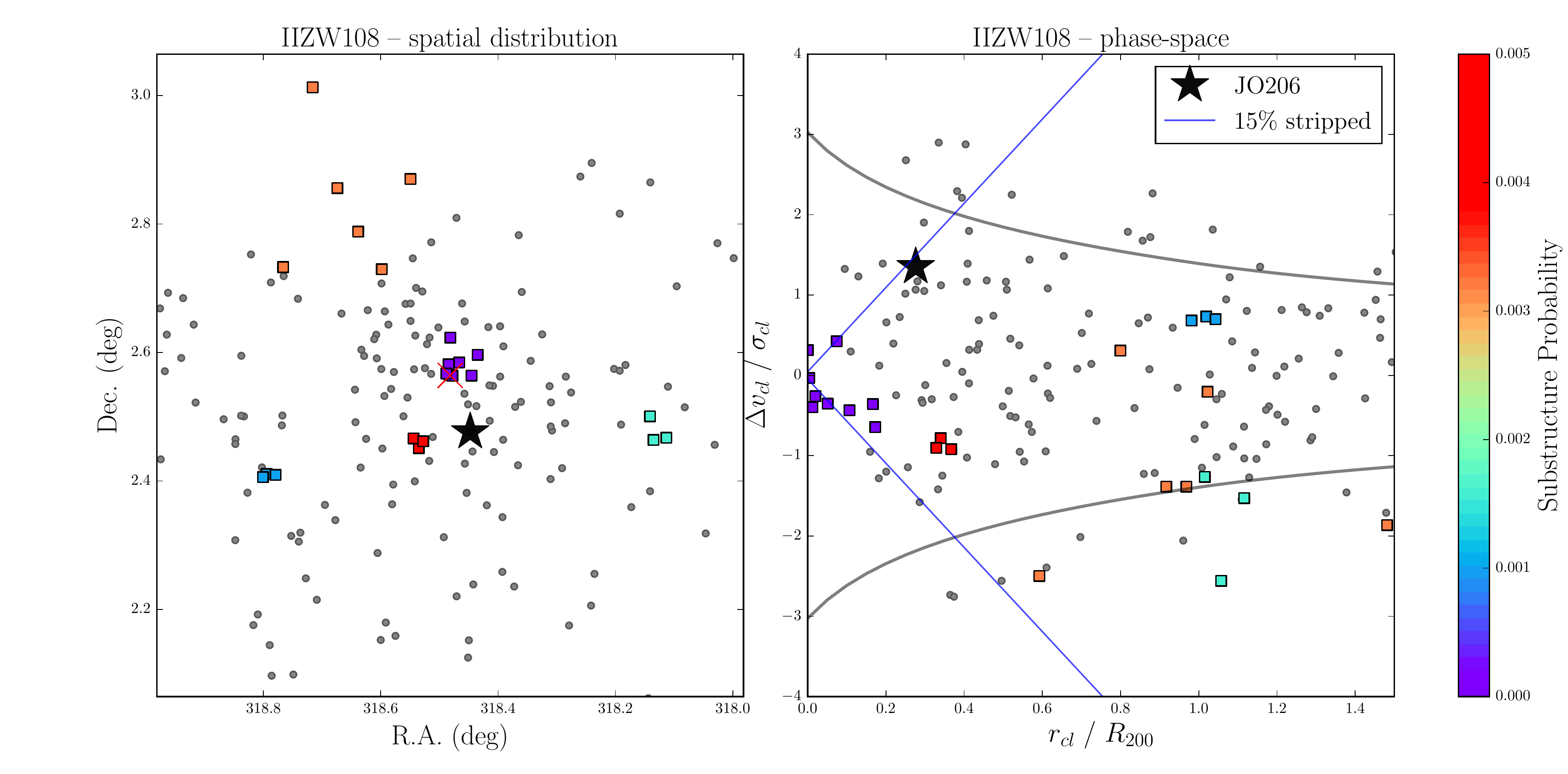}}
\caption{Left. Position on the sky of IIZW108 members (small
  black points) and non-members (small crosses), with large squares
  indicating the members of several substructures,  
color-coded according to the probability of these substructures to be
random fluctuations (i.e. values close to zero indicate highly
significant substructure detections, Biviano et al. in prep.).
JO206 is the dark big
  star. The red cross is the BCG. 
Right. Phase-space diagram with symbols as in the left panel, assuming
$\sigma = 545 \, \rm km/s$ and
$R_{200} = 1.17\rm \, Mpc$ (see text). Curves
show the escape velocity in a Navarro, Frenk \& White (1997) halo. The
blue line corresponds to 15\% of the total gas mass of JO206 
stripped due to ram-pressure by the ICM in a Virgo-like cluster (see 
text for details).
}
\end{figure*}


\section{Summary} 
GASP (GAs Stripping phenomena in galaxies with MUSE) is an ongoing ESO
Large Program with the MUSE spectrograph on the VLT.
This program started on October 1st 2015 and was allocated 120 hours
over four semesters to observe 114 galaxies 
to study the causes and the effects of gas removal
processes in galaxies.
GASP galaxies were homogeneously 
selected in clusters and in the field 
from the WINGS, OMEGAWINGS and PM2GC surveys, and belong to dark 
matter haloes with masses covering over four orders of magnitude. 

The main scientific drivers of GASP are a study of gas removal
processes in galaxies in different environments, their effects for the
star formation activity and quenching, the interplay between the gas
conditions and the AGN activity, and the stellar and metallicity
history of galaxies out to large radii prior to and in absence of gas removal.

The combination of large field-of-view,  high sensitivity, large 
wavelength coverage and good spatial and spectral resolution of MUSE 
allow 
to peer into the outskirts and 
surroundings of a large sample of galaxies with different masses and
environments.  The MUSE data is capable to reveal the rich physics of the 
ionized gas external to galaxies and the stars that formed
within it.

In this paper we have described the survey strategy and illustrated
the main steps of the scientific analysis, showing  their application
to JO206, a rather rare example of a massive ($9 \times 10^{10} M_{\odot}$) galaxy 
undergoing strong ram pressure stripping in a low mass cluster and forming $\sim 7 M_{\odot}
\, yr^{-1}$. Tails of stripped gas are
visible out to 90 kpc from the galaxy disk. The gas is ionized mosty
by in situ star formation, as new stars are formed in the tails.
The gas tails are characterized both by regions of diffuse emission
and bright knots, appearing to be giant HII regions and complexes,  that retain a
coherent rotation with the stars in the disk. The metallicity of the
gas varies over an order of magnitude from metal-rich regions on one
side of the disk, to very metal poor in some of the tails.
The galaxy hosts an AGN, that is responsible for $\sim 15$\% of
the $\rm H\alpha$ ionization.

The MUSE data reveal how the stripping and the star formation activity
and quenching have proceeded. The gas was stripped first from the
easternmost arm, where star formation stopped during the last few
$10^8$ yr. Star formation is still taking place in the disk, but
about a third of the total SFR (having excluded the AGN) takes place
outside of the main galaxy body, in the extraplanar gas and tails.
Assuming a Chabrier IMF, 1-2 $M_{\odot} \, yr^{-1}$ are formed outside of the galaxy disk, and go to increase
the intracluster light.

The first results shown in this and other papers of the series illustrate the power
of the MUSE data to provide an exquisite view of the
physical phenomena affecting the gas content of galaxies. 
GASP follow-up programs are undergoing to probe also the other gas phases and
obtain a multiwavelength view of this sample.
Ongoing APEX programs are yielding the CO amount in the disk and in the
tails. Approved JVLA observations will provide
the precious neutral gas information. Near-UV and far-UV data, as well
as complementary X-ray data, are being obtained for a subset of GASP
clusters with ASTROSAT (Subramaniam et al. 2016, Agrawal, P.C., 2006), while JWST will open
the possibility to obtain an unprecedented view of the $H_2$ gas.

\acknowledgments
We are grateful to the anonymous referee for his/her comments that  
significantly improved the presentation and the accuracy of the paper.  
Based on observations collected at the European Organisation for Astronomical Research in the Southern Hemisphere 
under ESO programme 196.B-0578. Based on observations taken with the
AAOmega spectrograph on the AAT, and the OmegaCAM camera on the VLT. 
This work made use of the KUBEVIZ software which is publicly available at
http://www.mpe.mpg.de/$\sim$dwilman/kubeviz/. We acknowledge financial
support from PRIN-INAF 2014. B.V. acknowledges the support from 
an Australian Research Council Discovery Early Career Researcher Award
(PD0028506). JF aknowledges financial support from UNAM-DGAPA-PAPIIT
IA104015 grant, M\'exico.This work was co-funded under the Marie Curie Actions of the European Commission (FP7-COFUND).
We warmly thank Matteo Fossati and Dave Wilman for their invaluable help
with KUBEVIZ, and Frederick Vogt for useful discussions and help for
optimizing {\it pyqz}. We are grateful to Joe Liske, Simon Driver and the whole MGC
collaboration for making their dataset easily available, and to Rosa
Calvi for her valuable work on the PM2GC.



\vspace{5mm}
\facilities{VLT(MUSE),VLT(OmegaCAM), AAT(AAOmega)}

\software{KUBEVIZ, ESOREX, SINOPSIS, IRAF, CLOUDY, pyqz, IDL, Python.}

\nocite{*}
\bibliography{references.bib}

\allauthors

\listofchanges

\end{document}